%% file: cuauvn_resub2.tex
\documentclass[twocolumn,letterpaper,aps,prc,superscriptaddress,nofootinbib,showpacs,showkeys,floatfix,dvipdfmx,longbibliography]{revtex4-1}

\usepackage{graphicx}	
\usepackage{xspace}	
\usepackage{hyperref}
\usepackage{url}
\usepackage{lineno}
\usepackage{bm}
\usepackage{amsmath}
\usepackage{epstopdf}
\usepackage[normalem]{ulem}

\newcommand{\pt}{\mbox{$p_T$}\xspace}
\newcommand{\mt}{\mbox{$m_T$}\xspace}
\newcommand{\Et}{\mbox{${\rm E}_T$}\xspace}
\newcommand{\sqsn}{\mbox{$\sqrt{s_{_{NN}}}$}\xspace}
\newcommand{\apt}{\mbox{$\langle p_T \rangle$}\xspace}
\newcommand{\apx}{\mbox{$\langle p_x \rangle$}\xspace}
\newcommand{\mpx}{\langle p_x \rangle}
\newcommand{\mpt}{\langle p_T \rangle}
\newcommand{\mean}[1]{\left\langle #1 \right\rangle}   

\usepackage{color}
\definecolor{orange}{cmyk}{0.,0.353,1.,0.}    

\begin{document}
%

\title{Azimuthal anisotropy in Cu+Au collisions at $\sqrt{s_{_{NN}}}$ = 200 GeV}

\input{authorlist_11282017.tex}

\date{\today}

\begin{abstract} 
The azimuthal anisotropic flow of identified and unidentified charged
particles has been systematically studied in Cu+Au collisions at
$\sqrt{s_{_{NN}}}$ = 200~GeV for harmonics $n=$~1--4 in the
pseudorapidity range $|\eta|<1$.  The directed flow in Cu+Au
collisions is compared with the rapidity-odd and, for the first time,
the rapidity-even components of charged particle directed flow in
Au+Au collisions at $\sqrt{s_{_{NN}}}$ = 200~GeV. The slope of the
directed flow pseudorapidity dependence in Cu+Au collisions is found
to be similar to that in Au+Au collisions, with the intercept shifted
toward positive pseudorapidity values, i.e., the Cu-going direction.
The mean transverse momentum projected onto the spectator plane,
$\langle p_x \rangle$, in Cu+Au collision also exhibits approximately
linear dependence on pseudorapidity with the intercept at about
$\eta\approx -0.4$ (shifted from zero in the Au-going direction),
closer to the rapidity of the Cu+Au system center-of-mass.  The
observed dependencies find natural explanation in a picture of the
directed flow originating partly due the ``tilted source'' and partly
due to the 
asymmetry in the initial density distribution.
A charge-dependence of $\langle p_x \rangle$ was also observed in
Cu+Au collisions, consistent with an effect of the initial electric
field created by charge difference of the spectator protons in two
colliding nuclei. The rapidity-even component of directed flow in
Au+Au collisions is close to that in Pb+Pb collisions at
$\sqrt{s_{_{NN}}}$ = 2.76 TeV, indicating a similar magnitude of
dipole-like fluctuations in the initial-state density distribution.
Higher harmonic flow in Cu+Au collisions exhibits similar trends to
those observed in Au+Au and Pb+Pb collisions and is qualitatively
reproduced by a viscous hydrodynamic model and a multi-phase transport
model. For all harmonics with $n\ge2$ we observe an approximate
scaling of $v_n$ with the number of constituent quarks; this scaling
works as well in Cu+Au collisions as it does in Au+Au collisions.
\end{abstract}

\maketitle

\setlength\linenumbersep{0.10cm}

\section{Introduction\label{sec:intro}}

The study of the azimuthal anisotropic flow in relativistic heavy-ion
collisions has been making valuable contributions to the exploration
of the properties of the hot and dense matter -- quark-gluon plasma
(QGP) -- created in such collisions.  Anisotropic flow is usually
characterized by the coefficients, $v_n$, in the Fourier expansion of
the particle azimuthal distribution measured relative to the so-called
flow symmetry planes: $dN/d\phi \propto 1 + 2 \sum_n v_n
\cos[n(\phi-\Psi_n)])$, where $\phi$ is the azimuthal angle of a
produced particle, and $\Psi_n$ is the azimuthal angle of the $n^{\rm
  th}$-harmonic flow plane.  The first harmonic (directed flow) and
second harmonic (elliptic flow) coefficients have been measured most
often and compared to the theoretical
models~\cite{Singha:2016mna,Heinz:2013th,Voloshin:2008dg}.
According to recent theoretical calculations, the higher harmonic flow
coefficients appear to provide additional and sometimes even stronger
constraints on the QGP models and on the initial conditions in
heavy-ion collisions~\cite{schenke,vnphenix}.

Elliptic flow, $v_2$, has been extensively studied both at the
Relativistic Heavy Ion Collider (RHIC) and the Large Hadron Collider
(LHC) energies.  For low transverse momentum ($\pt<2$ GeV/$c$),
$v_2(\pt)$ is well described by the viscous hydrodynamic models. A
comparison of the elliptic flow measurement to hydrodynamic model
calculations led to the finding that the QGP created in nuclear
collisions at RHIC and LHC energies has extremely small ratio of shear
viscosity to entropy density, $\eta/s$, and behaves as an almost ideal
liquid~\cite{Heinz:2013th,Voloshin:2008dg,schenke}.  The centrality
dependence of elliptic flow, and in particular flow fluctuations,
provided detailed information on the initial conditions and their
fluctuations.

While the experimental results on the elliptic flow are mostly
understood, there exists no single model that satisfactorily explains
the directed flow dependencies on centrality, collision energy, system
size, rapidity, transverse momentum, and even more, on the particle
type~\cite{Singha:2016mna}. This clearly indicates that an important
piece in our picture of ultrarelativistic collisions is still
missing. This could affect many conclusions made solely on the
elliptic flow measurements, as the initial conditions that would be
required for a satisfactory description of the directed flow, could
lead to stronger (or weaker) elliptic flow. Possible effects of that
have been mostly ignored so far in part due to complication of 3+1
hydrodynamical calculations compared to 2+1 calculations assuming
Bjorken scaling. The directed flow originates in the initial-state
spatial and momentum (initial collective velocity fields) asymmetries
in the transverse plane. The directed flow might be intimately related
to the vorticity in the system, and via that to the global
polarization of the system and to chirality flow -- two of the most
intriguing directions in current heavy ion
research~\cite{polBES,Kharzeev:2015znc}.

RHIC has been very successful in providing data on symmetric
collisions of approximately spherical nuclei such as Cu+Cu and Au+Au,
and non-spherical nuclei such as U+U, as well as asymmetric Cu+Au
collisions.  Since the anisotropic flow originates from the anisotropy
of the initial density distribution in the overlap region of the
colliding nuclei, these collisions provide important complementary
information on both the geometry and fluctuations in the initial
density distributions.  In particular, Cu+Au collisions are
characterized by a large asymmetry in the average initial density
distribution in the transverse plane, leading to significant $v_1$ and
$v_3$ flow coefficients even at midrapidity. Measurements of $v_1$ and
$v_3$ in Cu+Au collisions can be compared to the corresponding
measurements in symmetric collisions, where they can originate only in
density fluctuations, thus providing additional information on the
role of the initial density gradients.  Asymmetric collisions, with
their strong electric fields in the initial stages due to the charge
difference of spectator protons in the colliding nuclei, offer a
unique opportunity to study the electric conductivity of the created
matter and provide access to the time development of quark and
antiquark production~\cite{hirono,voronyuk,Voloshin:2016ppr,cuauv1_star}.

\begin{figure}[thb]
\begin{center}
\includegraphics[width=\linewidth]{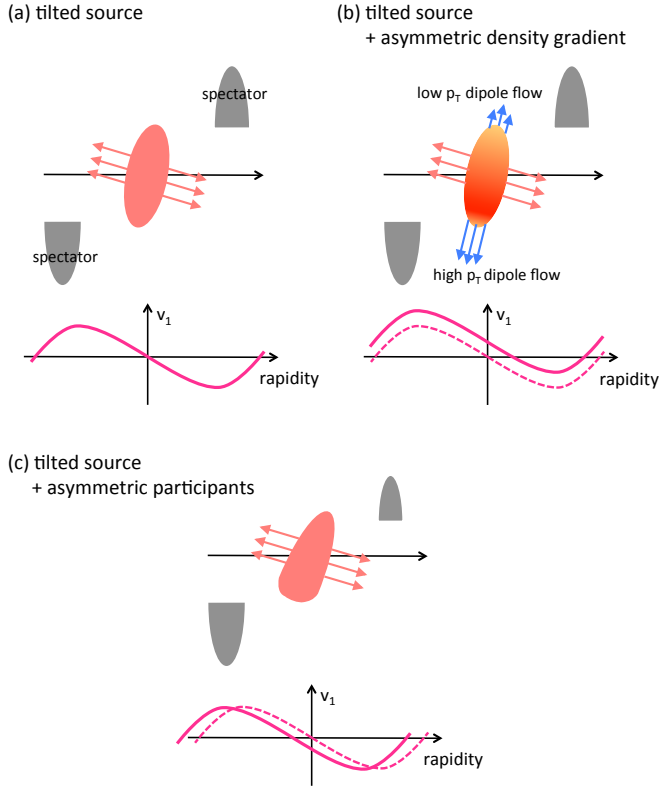}
\end{center}
\caption{(Color online) Cartoon illustrating different contributions
  to the directed flow and their effect on the (pseudo)rapidity
  dependence of mean $v_1$. Panel (a) shows the effect of the
  ``tilted source", while panels (b) and (c) include additional
  effects of asymmetric density distribution and asymmetry in number of
  participating nucleons.  In panels (b) and (c), the dashed lines represent
  the effect of the ``tilted source" only and the solid lines
  represent the two effects combined.}
\label{fig:tilt}
\end{figure}

In symmetric collisions, such as Au+Au, the directed flow measured
relative to the reaction plane (a plane defined by the impact
parameter vector and the beam direction) is an odd function of
(pseudo)rapidity. Note that while in symmetric collisions there exist an ambiguity/freedom 
in which of the nuclei is called a projectile and which a target, there is not any ambiguity in the results.  
The impact parameter is always defined as a vector in the transverse plane from the center of the target nucleus  
to the center of the projectile nucleus. The projectile velocity defines the positive $z$-direction, 
and, correspondingly, positive (pseudo)rapidity.
The directed flow measured relative to the reaction plane has a characteristic ``$\sim$''-shape, crossing
zero three times, with negative slope at midrapidity (for a review,
see~\cite{Singha:2016mna}), where the sign of the directed flow is
conventionally defined to be positive for projectile spectators at
forward rapidity.  The origin of such a dependence is not totally
clear. In hydrodynamic models, it is often produced through ``tilted''
source initial
conditions~\cite{Csernai:1999nf,Magas:2000jx,Bozek:2010bi}, as shown
in Fig.~\ref{fig:tilt}(a), with parameters of the tilt obtained from a
fit to the data~\cite{Bozek:2010bi,Becattini:2015ska}.  In a pure
``tilted source" scenario~\cite{Csernai:1999nf,Magas:2000jx},
$v_1(\pt)$ is a monotonic function of $\pt$ and the pseudorapidity
dependence of $\mpx(\eta) \equiv \mean{\pt \cos(\phi-\Psi_1)}$, where
$\mean{}$ means an average over particles in an event and then an
average over all events, can be directly related to that of
$v_1(\eta)$ (see Appendix).  In asymmetric collisions, as well as in
symmetric collisions away from midrapidity, the initial transverse
density distribution has dipole-like asymmetry.  This leads to an
additional contribution to anisotropic flow, interpreted either as
shadowing~\cite{Snellings:1999bt}, or due to the difference in
pressure gradients in different directions within the transverse
plane~\cite{v1_heinz}.  The first harmonic term, often called dipole
flow after a dipole-like density asymmetry, contributes to directed
flow.  The sign of the dipole flow contribution appears to be similar
to that of ``tilted source''.  However there exist a significant
difference between the two contributions -- the contribution to $\mpx$
from dipole flow is zero~\cite{TeaneyYan_2011}.  This fact can be used
to disentangle the relative contributions to directed flow from the
``tilted source'' and initial density asymmetries.  The condition
$\langle p_x \rangle^{\rm dipole}=0$ also leads to a characteristic
$v_1^{\rm dipole}(\pt)$ shape which crosses zero at $\pt \sim
\mpt$~\cite{TeaneyYan_2011}.  Higher $\pt$ particles tend to be
emitted in this direction, while lower $\pt$ particles are emitted in
the opposite direction to balance the momentum in the system.  The
sign of the average contribution to $v_1$ is determined by the low
$\pt$ particles.

The fluctuations in the initial density distribution, in particular
those leading to a dipole asymmetry in the transverse plane, lead to
non-zero directed flow, i.e. dipole flow, even at
midrapidity~\cite{TeaneyYan_2011}.  The direction (azimuthal angle) of
the initial dipole asymmetry, $\Psi_{1}^{\rm dipole}$, determines the
direction of flow.  The dipole flow angle $\Psi_{1}^{\rm dipole}$ can
be approximated by $\Psi_{1,3}=\arctan(\langle r^3\sin\phi
\rangle/\langle r^3 \cos\phi \rangle)+\pi$~\cite{TeaneyYan_2011} where
$r$ and $\phi$ are the polar coordinates of participants and a
weighted average is taken over the overlap region of two nuclei, with
the weight being the energy or entropy density.  The angle
$\Psi_{1,3}$ points in the direction of the largest density gradient.
Very schematically, the modification to $v_1(\eta)$ for a particular
fluctuation leading to positive dipole flow is shown in
Fig.~\ref{fig:tilt}(b).

The difference in the number of participating nucleons (quarks) in the
projectile and target nuclei also leads to the change in rapidity of
the ``fireball'' center-of-mass relative to that of nucleon-nucleon
system.  In symmetric collisions such a difference would be a
consequence of fluctuations in the number of participating nucleons
event-by-event~\cite{Csernai:2012mh}, while in asymmetric collisions
the position of the center-of-mass of participating nucleons will be
shifted on average, depending on centrality. In this case, one would
expect that the overall shape of $v_1(\eta)$ to be mostly unchanged,
but the entire $v_1(\eta$ curve to be shifted in the direction of
rapidity where more participants move, as schematically indicated in
Fig.~\ref{fig:tilt}(c).

Finally, we note that the 
dipole flow is found to be less sensitive to the shear viscosity over
entropy $\eta/s$~\cite{v1hydro} than $v_{2}$ and $v_{3}$, therefore it
provides a better constraint on the geometry and fluctuations of the
system in the initial state.

In Pb+Pb and Au+Au collisions the initial dipole-like asymmetry in the
density distribution at midrapidity is caused purely by the
fluctuations, while Cu+Au collisions have an intrinsic density
asymmetry due to the asymmetric size of colliding nuclei.  In addition
to the directed flow of the ``tilted source" (Fig.~\ref{fig:tilt}(a)),
one might expect the dipole flow to be produced by the asymmetric
density gradient (Fig.~\ref{fig:tilt}(b)) and the center-of-mass shift
in asymmetric collisions (Fig.~\ref{fig:tilt}(c)).  Therefore it is of
great interest to study the different components of directed flow in
Cu+Au collisions to improve our understanding of the role of gradients
in the initial density distributions and the hydrodynamic response to
such an initial state.

%

Experimentally, the directed flow is often studied with the first
harmonic event plane determined by the spectator
neutrons~\cite{v1smd,Abelev:2008jga,Adamczyk:2011aa}.  Recent
study~\cite{Voloshin:2016ppr} shows that in ultra-relativistic nuclear
collisions the spectators on average deflect outward from the center
of the collision, e.g projectile spectators deflect in the direction
of the impact parameter vector.  By combining the measurements
relative to the projectile, $\Psi_{\rm SP}^{p}$, and target,
$\Psi_{\rm SP}^{t}$, spectator planes, the ALICE Collaboration
reported the rapidity-odd and even components of directed flow in
Pb+Pb collisions at \sqsn = 2.76 TeV~\cite{aliceV1}:
\begin{eqnarray}
v_1 &=& v_1^{\rm odd} + v_1^{\rm even},\\ 
v_{1}^{\rm odd} &=& (v_{1}\{\Psi_{\rm SP}^{p}\} - v_{1}\{\Psi_{\rm SP}^{t}\}
)/2, \label{eq:v1odd} \\ 
v_{1}^{\rm even} &=& ( v_{1}\{\Psi_{\rm
  SP}^{p}\} + v_{1}\{\Psi_{\rm SP}^{t}\} )/2. \label{eq:v1even},
\end{eqnarray}
where the ``even'' component might originate in the fluctuation of the
initial density.  Note that the
``projectile'' nucleus defines the forward direction and
$\langle\cos(\Psi_{\rm SP}^{p}-\Psi_{\rm SP}^{t})\rangle<0$.  Since
the target spectator plane $\Psi_{\rm SP}^t$ points in the opposite
direction to $\Psi_{\rm SP}^{p}$, in the ALICE paper~\cite{aliceV1},
directed flow relative to the target spectator plane was defined as
$v_1\{\Psi_{\rm SP}^{t}\} = -\langle\cos(\phi-\Psi_{\rm
  SP}^{t})\rangle$, resulting in Eqs.~\eqref{eq:v1odd} and
\eqref{eq:v1even} having the opposite sign convention from
Ref.~\cite{aliceV1}.

A finite $v_1^{\rm even}$ was observed in Pb+Pb collisions with little
if any rapidity dependence~\cite{aliceV1}. It is believed that the
origin of this component is in finite correlations between the
direction of spectator plane and the direction of the initial dipole
asymmetry at midrapidity. Such a correlation is expected to be weak,
$\langle\cos(\Psi_{\rm SP}^{p}-\Psi_{1,3})\rangle \ll 1$, which would
explain the small magnitude of $v_{1}^{\rm even}$ of the order of a
few per mil.  The $v_1^{\rm dipole}$ can be measured via two-particle
correlation ($v_1^{\rm dipole}$ relative to participant
plane)~\cite{ATLAS:2012at,Aamodt:2011by} taking into account the
momentum conservation effect which requires model-dependent treatment.
The $v_1^{\rm dipole}$ measured using two-particle
correlation~\cite{ATLAS:2012at} shows $\sim$40 times larger magnitude
than $v_1^{\rm even}$ measured with spectator planes. This difference
can be explained by the weak correlation of $\langle\cos(\Psi_{\rm
  SP}^{p}-\Psi_{1,3})\rangle$ as discussed in Ref.~\cite{aliceV1}.

%
\begin{figure}[thb]
\begin{center}
\includegraphics[width=\linewidth]{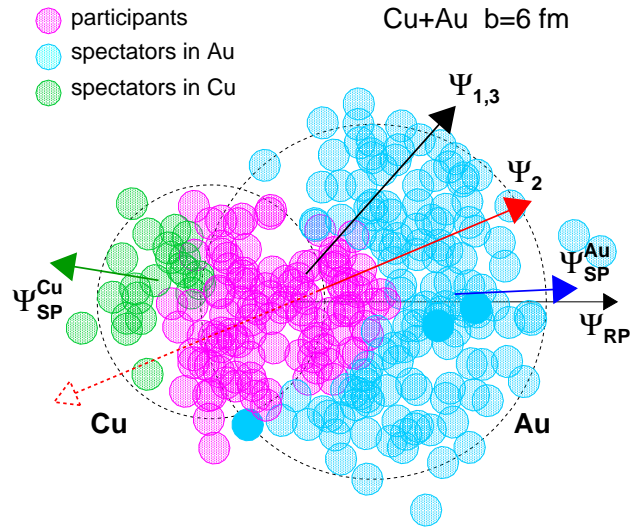} 
\end{center}
\caption{(Color online) Cartoon of Cu+Au collision indicating
  different event planes used in the analysis. Note that $\Psi_2$ and
  $\Psi_2+\pi$ define the same plane.}
\label{fig:planes}
\end{figure}

Following a similar approach to that of ALICE Collaboration, we study
directed flow in midrapidity region relative to the target (Au) and
projectile (Cu) spectator planes (see Fig.~\ref{fig:planes}). We
identify two components of the directed flow: the one determined by
the directed flow relative to the (true) reaction plane, $\Psi_{\rm
  RP}$, and the component due to the initial density fluctuations. The
first component is similar to the ``odd'' component in symmetric
collisions, but in Cu+Au collisions it also includes a contribution
due to non-zero average dipole-like asymmetry in the initial density
distribution. The second component, due to the initial density
fluctuations, is similar to the ``even'' component in the ALICE
analysis. In addition to the results obtained from correlations to the
spectator planes, we also present the results from 3-particle
correlations~\cite{Voloshin:2008dg,v2star,v1smd}, $v_1\{3\}$, which are
interpreted as projection of the directed flow onto the second
harmonic event plane, $\Psi_2$, that is defined by participants. See
the schematic view of a collision with different event planes
identified in Fig.~\ref{fig:planes}.  Model
calculations~\cite{TeaneyYan_2011} suggest that the dipole flow might
be correlated more strongly with $\Psi_2$ (second harmonic participant
event plane) than with the spectator plane (which is very close to the
reaction plane), and thus one can expect that the dipole flow
contribution to $v_1\{3\}$ might be slightly larger than that with the
spectator plane.

Elliptic and higher harmonic flow measurements in asymmetric
collisions are also extremely interesting. While in symmetric
collisions, the odd harmonics originate from the initial density
fluctuations~\cite{Alver:2010gr}, in asymmetric collisions the
intrinsic geometrical asymmetry in the initial state may lead to
significant odd components of the flow.  Thus the measurements of
higher harmonic flow as well as the directed flow in Cu+Au collisions
provide an opportunity to study the interplay of the two effects and
provide additional constraints on hydrodynamic models.

A quark number scaling was observed for the elliptic
flow~\cite{Voloshin:2008dg,v2ket_star2007,v2ket_phenix2007},
suggesting collective behavior at a partonic level. Recently PHENIX
reported that the quark number scaling also works for higher harmonic
flow in Au+Au collisions at \sqsn = 200 GeV~\cite{pidvn_phenix} by
considering the order of the harmonics in the scaling rule, although
the interpretation is still under discussion.  It is very interesting
to study if such a scaling is also held in asymmetric collisions
having a potentially different origin for the odd component of the
higher harmonic flow.

In this paper we present the measurements of the higher harmonic (up
to $n=4$) anisotropic flow of unidentified and identified charged
particles in Cu+Au collisions.  Results from Cu+Au collisions are
compared with those from Au+Au collisions, as well as with
hydrodynamic and transport models.  We discuss the quark number
scaling for $v_2$, $v_3$, and $v_4$ of charged pions, charged kaons,
and (anti)protons. Compared to the previous measurements, 
a better accuracy of $v_3$ results and new data on
$v_4$ provide a more detailed view on the scaling properties of
anisotropic flow in asymmetric collisions and the
physics behind it.

This paper is organized as follows: Section \ref{sec:exp} provides a
brief explanation of the experimental setup. The details of data
reduction and analysis method are described in Sec.~\ref{sec:ana}.
Results for the directed flow are presented in Sec.~\ref{sec:v1} and
results for higher harmonic flow are presented in Sec.~\ref{sec:vn}.
For charged particles, we compare our results to theoretical models.
For the higher harmonic flow of identified particles, we also discuss
the number of constituent quark (NCQ) scaling.
Section~\ref{sec:summary} summarizes the results and findings.

%
\section{Experimental Setup\label{sec:exp}}

The STAR detector system is composed of central detectors
performing tracking and particle identification, and trigger detectors
located at the forward and backward directions. The Zero Degree
Calorimeters (ZDC)~\cite{zdc} and the Vertex Position Detector
(VPD)~\cite{vpd} are used to determine the minimum-bias trigger. The
ZDCs are located at forward and backward angles of $|\eta|>6.3$ and
measure the energy deposit of spectator neutrons. The VPD consists of
two identical detectors surrounding the beam pipe and covering the
pseudorapidity range of $4.24<|\eta|<5.1$. The VPD provides the start
time of the collision and the position of the collision vertex along
the beam direction.

The Time Projection Chamber (TPC)~\cite{tpc} is used for the tracking
of charged particles. It covers the full azimuth and has an active
pseudorapidity range of $|\eta|<1$. The TPC is also used for particle
identification via specific ionization energy loss, $dE/dx$. Particle
identification also utilizes the Time-Of-Flight detector
(TOF)~\cite{tof}. The TOF consists of multigap resistive plate
chambers and covers the full azimuth and has a pseudorapidity range of
$|\eta|<0.9$. The timing resolution of the TOF system with the start
time from the VPD is $\sim$100~ps.

%
\section{Data Analysis\label{sec:ana}}

The analysis is based on the minimum-bias data for Cu+Au collisions at
\sqsn = 200~GeV collected in 2012 and Au+Au collisions at \sqsn = 200
GeV collected in 2010. The collision vertex was required to be within
$\pm$30~cm from the center of the TPC in the beam direction.
Additionally, the difference between the two $z$-vertex positions
determined by TPC and VPD was required to be less than
$\pm$3~cm to reduce the beam-induced background (pileup). The vertex position
in the transverse plane was required to be within 2~cm from the
beam center. These criteria select forty-four million minimum-bias
triggered events for Cu+Au collisions and ninety-five million minimum-bias 
triggered events for Au+Au collisions. Centrality was defined
based on the measured charged particle multiplicity within
$|\eta|<0.5$ and a Monte Carlo Glauber simulation in the same way as
in previous studies~\cite{BESv2}. The effect of the trigger efficiency
was taken into account in the results by appropriate weights for both
Cu+Au and Au+Au collisions.

In the following subsections, the details of analysis are
described. Analysis procedures are basically the same as in previous STAR
publications~\cite{v2star,bes_pidv2}. The only difference in the analysis
between asymmetric and symmetric collisions is the way to evaluate the
resolution of the event plane because one cannot assume equal
subevents in forward and backward rapidities (two subevent method) in
asymmetric collisions, as explained in Sec.~\ref{sec:ep} and
\ref{sec:flow}.

%
\subsection{Track selection and particle identification\label{sec:track}}

Good quality charged tracks were selected based on the TPC hit
information as follows. The number of hit points used in track
reconstruction was required to be greater than 14, with the maximum
possible number of hit points of 45. The ratio of the number of hit
points to the maximum possible for that track was required to be
larger than~0.52.  These requirements ensure better momentum
resolution and allow to avoid track splitting and merging effects.
The track distance of closest approach to the primary vertex (DCA),
was required to be less than 3~cm to reduce contributions from
secondary decay particles.  The tracks within $0.15<\pt<5$~GeV/$c$ and
$|\eta|<1$ were analyzed in this study.

%

Particle identification was performed using the TPC and TOF
information as mentioned above. For the TPC, the particles were
identified based on the $dE/dx$ distribution normalized by the
expected energy loss given by the Bichsel function~\cite{bichsel},
expressed as $n\sigma^{\rm TPC} = \log[(dE/dx)^{\rm meas}/(dE/dx)^{\rm
    exp}]/\delta_{dE/dx}$, where $\delta_{dE/dx}$ is the $dE/dx$
resolution.  The distribution of $n\sigma^{\rm TPC}$ is nearly
Gaussian for a given momentum and is calibrated to be centered at zero
with a width of unity for each particle
species~\cite{Xu:2008th,Adamczyk:2013gw}.  $\pi^{+}(\pi^{-})$,
$K^{+}(K^{-})$, and $p(\bar{p})$ samples were obtained by requiring
$|n\sigma^{\rm TPC}|<2$ for particles of interest and $|n\sigma^{\rm
  TPC}|>2$ for other particle species. To increase the purity of the
kaon and proton samples, we applied the more stringent pion rejection
requirement $|n\sigma^{\rm TPC}|>3$.  When the track has hit
information from the TOF, the squared mass ($m^2$) can be calculated
from the momentum, the time of flight, and the path length of the
particle. The $\pi^{+}(\pi^{-})$, $K^{+}(K^{-})$, and $p(\bar{p})$
were selected from a 2$\sigma$ window relative to their peaks in the
$m^2$ distribution. Additionally the selected particles were required
to be away from the $m^2$ peak for other particles. When the TOF
information was used in the particle identification, the TPC selection
criterion was relaxed to $|n\sigma^{\rm TPC}|<3$ for the particle of
interest.
The purity of selected samples drops down to $\sim$90\% at higher \pt.
However we found that the variation of particle selection cuts does
not affect the results beyond the uncertainties as described in
Sec.~\ref{sec:sys}.

%
\subsection{Event plane determination}\label{sec:ep}

The event plane angles were reconstructed based on the following
equations~\cite{Voloshin:2008dg}:
\begin{eqnarray}
n\Psi_{n}^{\rm obs} &=& \tan^{-1}\left(\frac{Q_{n,y}}{Q_{n,x}}\right), 
\label{eq:psin}\\ 
Q_{n,x} &=& \sum_{i} w_{i} \cos(n\phi_{i}), 
\label{eq:qx}
\\ Q_{n,y} &=& \sum_{i} w_{i} \sin(n\phi_{i}) , 
\label{eq:qy}
\end{eqnarray}
where $\phi_{i}$ is the azimuthal angle of the charged track and
$w_{i}$ is the \pt weight (used only for the event plane determined in
the TPC). The $\Psi_{n}^{\rm obs}$ is an estimated n$^{\rm th}$-order
event plane and $Q_{n,x(y)}$ is referred to as the flow
vector. Corrections for the detector acceptance were applied following
Ref.~\cite{EPcorrection}.  The tracks measured in the TPC acceptance
were divided into three subevents ($-1<\eta<-0.4$, $|\eta|<0.2$, and
$0.4<\eta<1$). The track selection criteria mentioned above were
applied but only tracks with $p_{T}<2$ GeV/$c$ were used for the event
plane reconstruction.

The Beam-Beam Counters (BBC)~\cite{bbc} and the Endcap-Electromagnetic
Calorimeter (EEMC)~\cite{eemc} were also used for the event plane
determination in addition to the TPC. The BBCs are located at forward
and backward angles ($3.3<|\eta|<5$) and consist of scintillator
tiles. When using the BBCs for the event plane determination, the
azimuthal angle of the center of each tile was used for $\phi_{i}$ in
Eqs.~\eqref{eq:qx} and \eqref{eq:qy}. and the ADC value in that tile
was used as the weight, $w_i$. The EEMC covers the pseudorapidity
range of $1.086<\eta<2$ and consists of 720 towers ($60\times12$ in
$\phi-\eta$ plane). When using the EEMC for the event plane
determination, the azimuthal angle of each tower center was used as
$\phi_{i}$, and the transverse energy, \Et, was used as $w_i$. If \Et
exceeded 2~GeV, a constant value of 2 was used as the weight.

For the first-order event plane, the ZDCs with Shower Maximum
Detectors (SMD)~\cite{v1smd} were used. Each SMD is composed of two
planes with scintillator strips aligned with the $x$ or $y$ directions
and sandwiched between the ZDC modules. Therefore, the SMD measures
the centroid of the hadronic shower caused by the interaction between
spectator neutrons and the ZDC.  The x and y positions of the shower
centroid was calculated for each ZDC-SMD on the event-by-event basis as
follows:
\begin{eqnarray}
\langle X \rangle &=& \frac{\sum_i X_i \cdot w_{X_i}}{\sum_i w_{X_i}} \label{eq:qxsmd} \\
\langle Y \rangle &=& \frac{\sum_i Y_i \cdot w_{Y_i}}{\sum_i w_{Y_i}} \label{eq:qysmd}
\end{eqnarray}
where $X_i(Y_i)$ denotes the position of a vertical (horizontal)
scintillator strip in the SMD and $w_{X_i}(w_{Y_i})$ denotes the ADC signal
measured in each strip. Then the first-order event plane was
determined as $\Psi_1 = \tan^{-1} (\langle Y\rangle/\langle
X\rangle)$.  The angle determined by the target spectators points into the
opposite direction (+$\pi$) to that of the projectile spectator plane, then
the combined event plane of ZDC-SMD east and west can be obtained by
summing Eqs.~\eqref{eq:qxsmd} and \eqref{eq:qysmd} from each ZDC-SMDs
flipping the sign for one of them.

%
\begin{figure}[thb]
\begin{center}
\includegraphics[width=\linewidth,trim=0 30 0 0]{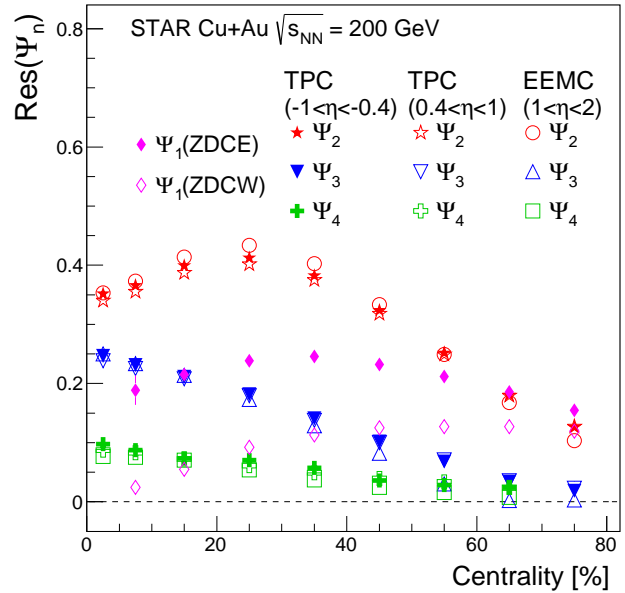}
\end{center}
\caption{(Color online) Event plane resolutions as a function of
  centrality in Cu+Au collisions at \sqsn = 200 GeV.}
\label{fig:epres}
\end{figure}
The event plane resolution defined as ${\rm
  Res}(\Psi_n)=\langle\cos(\Psi_n-\Psi_n^{\rm obs})\rangle$ was
estimated by the three-subevent method~\cite{TwoSub}.  Here
$\Psi_n^{\rm obs}$ denotes the azimuthal angle of a measured
(``observed'') event plane.  For the first-order event plane, either
BBC in the west (BBCW) or east (BBCE) sides was used as a third
subevent along with the two ZDCs.  For higher harmonic event plane,
three subevents from TPC were used.  In the case of using the EEMC,
one of the TPC subevents was replaced with EEMC subevent.  In Au+Au
collisions, both the two-subevent and the three-subevent methods were
used.  The results are reported using the reaction plane resolution
from the two-subevent method, with the difference in results between
the two methods included in the systematic
uncertainty. Figure~\ref{fig:epres} shows the estimated event plane
resolution, ${\rm Res(\Psi_{n})}=\langle\cos(\Psi_n-\Psi_n^{\rm
  obs})\rangle$ ($2\leq n\leq4$), for TPC and EEMC, and ${\rm
  Res(\Psi_{1})}$ for ZDC-SMD in Cu+Au collisions. Note that the
forward direction or the west side (ZDCW and BBCW) is the Cu-going
direction.  The resolution of $\Psi_{1}$ with ZDC-SMD in Au+Au
collisions can be found in Ref.~\cite{Wang:2007kz}.  Results for wide
centrality bins in this study were obtained by taking averages of
results measured with 10\% step centrality bins.

%
\subsection{Flow measurements}\label{sec:flow}
Azimuthal anisotropy was measured with the event plane method using
the following equation:
%
\begin{eqnarray}
v_{n} = \frac{\langle \cos[n(\phi-\Psi_{n}^{\rm obs})] \rangle}{\rm Res(\Psi_{n})},
\end{eqnarray}
where $\langle\,\rangle$ means an average over particles in an event,
followed by the averaging over all events.  We study $v_n$ as a
function of \pt\ for different centralities, as well as the
(pseudo)rapidity dependence of $v_1$.  For the event plane determined
by TPC, the $v_{n}$ of charged particles were measured using an
$\eta$-gap of 0.4 from the subevent used for the event plane
determination, i.e. particles of interest were taken from $-1<\eta<0$
($0<\eta<1$) when using the event plane determined in the subevent
from the forward (backward) rapidity. The results from these two
subevents are found to be consistent and the average of the two
measurements is used as the final result.

%
Directed flow can be also measured by the three-point correlator with the
use of the second harmonic event plane~\cite{v2star}:
\begin{eqnarray}
v_{1}\{3\} = \frac{\langle \cos(\phi+\Psi_{1}^{\rm obs}-2\Psi_{2}^{\rm obs}) \rangle}{\rm Res(\Psi_{1}) \times Res(\Psi_{2})},
\label{eq:v1_3}
\end{eqnarray}
where $\Psi_{1}^{\rm obs}$ and $\Psi_{2}^{\rm obs}$ were taken from
different subevents and $\phi$ is the azimuthal angle of particles of
interest in the rapidity region different from those subevents to
avoid self-correlation. In our analysis, $\Psi_1^{\rm obs}$ was taken
from the east BBC and $\Psi_2^{\rm obs}$ from either the TPC or EEMC
subevents. The results for $v_{1}\{3\}$ obtained with TPC subevents
from the backward and forward rapidities are statistically consistent
in the overlapping region, and were further combined to cover the same
$\eta$ range for particles of interest as used in the event plane
method. The difference between results obtained from TPC or EEMC
subevents was taken into account as a systematic uncertainty. Note
that Eq.~\eqref{eq:v1_3} was calculated without any spectator
information, and thus provides information on the directed flow
projected onto the second harmonic participant plane.

For the higher harmonic flow measurements, the scalar product
method~\cite{Adam:2016nfo,SP,vnSP} was tested for comparison with the
event plane method. The scalar product method is equivalent to the
two-particle correlation method with corresponding $\eta$ gap between
two particles and particle of interest. Three subevents were used to
calculate the flow coefficients based on the following equation:
\begin{eqnarray}\label{eq:vnsp}
v_{n} = \frac{\langle \bm{u} \cdot \bm{Q}_{n}^{A}/N^{A}
  \rangle}{\sqrt{ \cfrac{ \langle\bm{Q}_{n}^{B}/N^{B} \cdot
      \bm{Q}_{n}^{C}/N^{C}\rangle }{ \langle\bm{Q}_{n}^{A}/N^{A} \cdot
      \bm{Q}_{n}^{B}/N^{B}\rangle \langle\bm{Q}_{n}^{C}/N^{C} \cdot
      \bm{Q}_{n}^{A}/N^{A}\rangle } }},
\end{eqnarray}
where $\bm{Q}_{n}$ is the flow vector defined in Eqs.~\eqref{eq:qx}
and \eqref{eq:qy} and the superscripts $A$, $B$, and $C$ denote
different subevents with a finite rapidity gap from the other
subevent. The subevents were taken from TPC and/or EEMC.  We denote by
$\bm{u}$ a unit vector in the direction of the particle transverse
momentum; $N$ denotes the sum of weights used for
reconstructing the flow vectors in each subevent.

The tracking efficiency was accounted for in \pt-integrated
observables, although the effect of that is much smaller than other
systematic uncertainties discussed below.

\subsection{Systematic uncertainties}\label{sec:sys}

The systematic uncertainties were estimated by varying the track
quality cuts described in~\ref{sec:track} and by varying collision
$z$-vertex cut. The effect of the track quality cuts becomes largest
at low \pt in central collisions and was found to be $<$4\% for $v_2$,
$<$6\% for $v_3$, and $<$8\% for $v_4$. The effect of the $z$-vertex
cut is $<$1\%. For identified particles, the effect of particle
identification purity was also considered. The effect for charged
pions is $<$1\% in $v_2$ and $v_3$ and $<$3\% in $v_4$. The effects
for charged kaons and (anti)protons are $<$3\% in $v_2$, $<$5\% in
$v_3$, and $<$10\% in $v_4$. The combined estimated uncertainty was
found to be \pt-uncorrelated; namely all data points do not move in
the same direction over \pt, and was assigned as a point-by-point
systematic uncertainty.
\begin{figure*}[bth]
\begin{center}
\includegraphics[width=0.85\linewidth,trim=0 12 0 10]{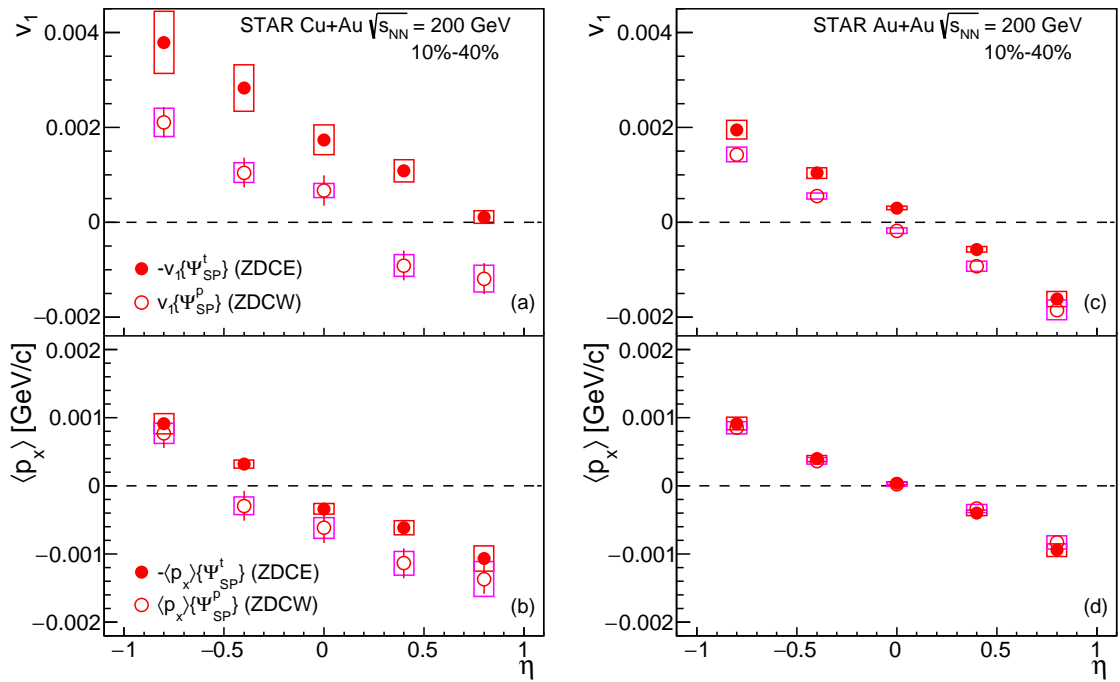}
\caption{(Color online) Directed flow of charged particles measured
  with respect to the target (ZDCE) and projectile (ZDCW) spectator
  planes and the mean transverse momentum projected onto the spectator
  planes, as a function of $\eta$ for $0.15<\pt<5$ GeV/$c$ in
  10\%-40\% centrality for Cu+Au (a,b) and Au+Au (c,d) collisions at
  \sqsn = 200 GeV. Open boxes show the systematic uncertainties.  Note
  that the directed flow obtained with the target spectator plane
  ($v_{1}\{\Psi_{\rm SP}^{t}\}$) is shown with opposite sign.}
\label{fig:v1eta_vsZDC}
\end{center}
\end{figure*}

Along with the TPC event plane, the event plane determined by the EEMC
was used for the $v_n$ ($n\geq2$) measurements and the difference in
$v_n$ obtained with the two methods was included in the systematic
uncertainty. The latter was found to be \pt-correlated: it was $<$2\%
($<$10\%) for $v_2$ and $v_3$ ($v_4$) in central collisions, and
increased up to $\sim$5\% (16\%) for $v_2$ ($v_3$ and $v_4$) in
peripheral collisions. For $v_1$, the details of the systematic
uncertainty estimation can be found in our previous
study~\cite{cuauv1_star}. As mentioned before, $v_{1}\{3\}$ was
measured without the spectator information, but one can also use the
ZDCs for $\Psi_{1}$ in Eq.~\eqref{eq:v1_3} for a cross check. We found
that $v_{1}\{3\}$ measured using the ZDCs was consistent with
$v_{1}\{3\}$ measured using the BBC within the uncertainties.

%
%
\section{Directed flow\label{sec:v1}}
\subsection{Directed flow of unidentified hadrons\label{sec:chv1}}
The top panels (a,c) of Fig.~\ref{fig:v1eta_vsZDC} present the
directed flow, $v_1$, of charged particles as a function of the
pseudorapidity with respect to the target and projectile spectator
planes in Cu+Au and Au+Au collisions at \sqsn = 200~GeV. It is taken
into account that the projectile spectators deflect on average along
the impact parameter vector (a vector from the center of the target to
the center of the projectile, taken in this analysis to be Cu
nucleus)~\cite{Voloshin:2016ppr}.
The sign of $v_1$ measured with respect to the target spectator plane
has been reversed. In both systems, a finite difference can be seen
between $v_1$ measured with respect to each spectator plane. This
indicates the existence of a fluctuation component (rapidity-even for
symmetric collisions) of $v_1$ in both symmetric and asymmetric
collision systems.

The notion of ``odd" and ``even" $v_1$ components can be justified
only for symmetric collisions. Therefore, the following definitions
are used for Cu+Au collisions:
\begin{eqnarray}
v_{1}^{\rm conv} &=& ( v_{1}\{\Psi_{\rm SP}^{p}\} - v_{1}\{\Psi_{\rm
  SP}^{t}\} )/2 \label{eq:v1conv} \\ v_{1}^{\rm fluc} &=& (
v_{1}\{\Psi_{\rm SP}^{p}\} + v_{1}\{\Psi_{\rm SP}^{t}\}
)/2,
 \label{eq:v1fluc}
\end{eqnarray}
where ``projectile" (Cu) spectators go into the forward direction.
The term $v_{1}^{\rm conv}$ and $v_{1}^{\rm fluc}$ denotes
``conventional" and ``fluctuation'' components of directed flow,
respectively.  Note that the right-hand side of Eq.~\eqref{eq:v1conv}
and Eq.~\eqref{eq:v1fluc} represents the same definitions as
Eq.~\eqref{eq:v1odd} and Eq.~\eqref{eq:v1even}.

The mean transverse momentum projected onto the spectator plane defined as
\begin{eqnarray}
\langle p_x \rangle 
= \frac{\langle\pt\cos(\phi-\Psi_{1}^{\rm obs})\rangle}{{\rm Res}(\Psi_{1})},
\end{eqnarray}
is also shown in the bottom panels (b,d) of
Fig.~\ref{fig:v1eta_vsZDC}.  There seems to be small difference
between results with two spectator planes in Cu+Au but not in Au+Au.
The terms ``conv (odd)" and ``fluc (even)" are also used for $\mpx$ in
the following discussion, with analogous definitions to
Eqs.~\eqref{eq:v1conv} and \eqref{eq:v1fluc}.
%
\begin{figure*}[!thb]
\begin{center}
\includegraphics[width=0.85\linewidth,trim=0 14 0 10]{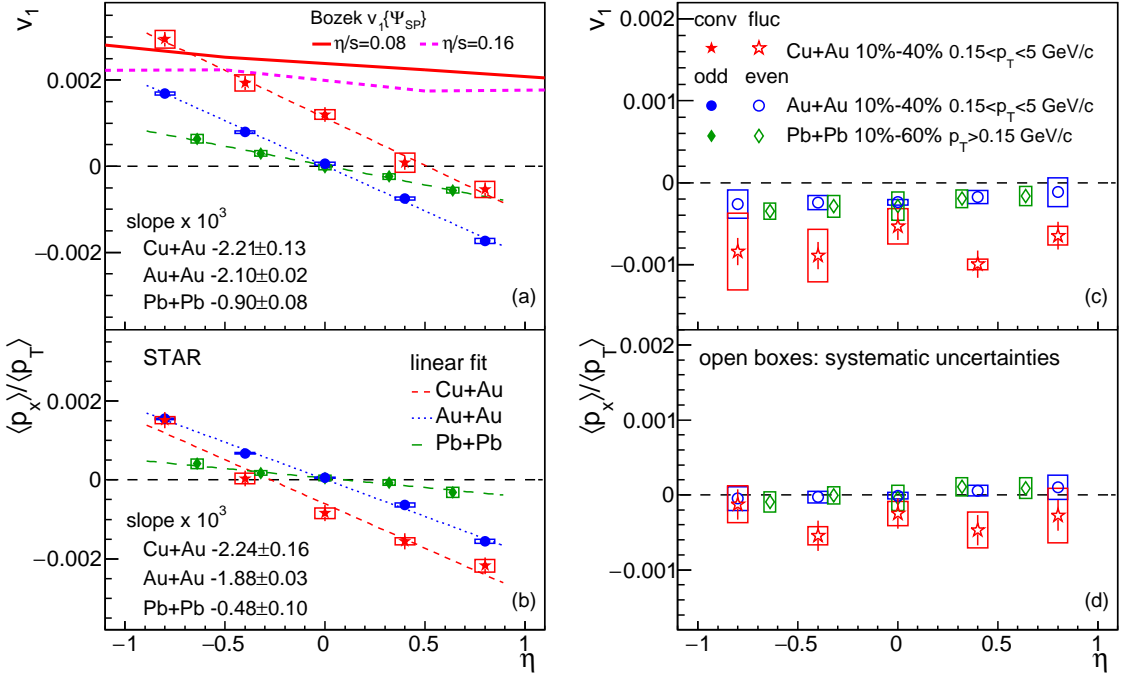}
\end{center}
\caption{(Color online) Charged particle ``conventional" (left) and
  ``fluctuation" (right) components of directed flow $v_1$ and
  momentum shift $\apx/\apt$ as a function of $\eta$ in 10\%-40\%
  centrality for Cu+Au and Au+Au collisions at \sqsn = 200 GeV, and
  Pb+Pb collisions at \sqsn = 2.76 TeV~\cite{aliceV1}.  Thick solid
  and dashed lines show the hydrodynamic model calculations with
  $\eta/s$=0.08 and 0.16, respectively, for Cu+Au
  collisions~\cite{bozek}.  Thin lines in the left panel show a linear
  fit to the data. Open boxes represent systematic uncertainties.}
\label{fig:v1px_eta}
\end{figure*}

The top panels of Fig.~\ref{fig:v1px_eta} present the pseudorapidity
dependence of $v_{1}^{\rm odd (conv)}$ and $v_{1}^{\rm even (fluc)}$,
defined according to Eqs.~\eqref{eq:v1odd}, \eqref{eq:v1even},
\eqref{eq:v1conv}, and \eqref{eq:v1fluc}.  The $\mpx$ normalized by
the mean \pt is also shown in the bottom panels.  The lines represent
linear fits to guide the eye. The conventional component of directed
flow, $v_1^{\rm conv}$, in Cu+Au has a similar slope to $v_1^{\rm
  odd}$ in Au+Au, with the intercept shifted to the forward direction.
The mean transverse momentum component $\langle p_{x}^{\rm conv}
\rangle$ in Cu+Au might deviate from linear dependence (observed in
Au+Au) with the slope slightly increasing at backward rapidities. This
trend in $\langle p_{x}^{\rm conv}\rangle$ might reflect the momentum
balance between particles produced in the forward and backward
hemispheres -- in Cu+Au collisions more charged particles are produced
in the Au-going direction, and therefore the particles at forward
rapidity need to have a larger $p_x$ on average to compensate the
asymmetric multiplicity distribution over $\eta$.  Results from Pb+Pb
collisions at \sqsn = 2.76 TeV measured by the ALICE
experiment~\cite{aliceV1} are also shown in Fig.~\ref{fig:v1px_eta}.
The slope of $v_1^{\rm odd}$ in Pb+Pb collisions is about 3 times
smaller than that in Au+Au collisions.  This trend, i.e. the energy
dependence of the $v_1$ slope, is consistent with that observed in the
RHIC Beam Energy Scan~\cite{BES_pv1}. Calculations from an
event-by-event hydrodynamic model with two different values of
$\eta/s$ ($\eta/s$ = 0.08 and 0.16) for Cu+Au collisions~\cite{bozek}
are also compared to the data. Despite the model's successful
description of elliptic flow and triangular flow (see
Section~\ref{sec:vn}), it cannot reproduce either the magnitude of the
directed flow nor its pseudorapidity dependence.

The even component of directed flow, $v_1^{\rm even}$, in Au+Au does
(Fig.~\ref{fig:v1px_eta}(c)) not depend on pseudorapidity (within
error bars) and is very similar in magnitude to $v_1^{\rm even}$ in
Pb+Pb collision at LHC energies.  The $\langle p_x^{\rm even}\rangle$
in both Au+Au and Pb+Pb collisions is consistent with zero, which
indicates zero net transverse momentum in the systems. This agrees
with the expectation that the even component of $v_1$ originates from
event-by-event fluctuations of the initial density. The magnitude of
$v_1^{\rm fluc}$ in Cu+Au is larger than that of $v_1^{\rm even}$ in
Au+Au. This would be due either to larger initial density fluctuations
in Cu+Au collisions or to stronger correlations between the spectator
and dipole fluctuation planes.
%
\begin{figure*}[thb]
\begin{center}
\includegraphics[width=0.42\linewidth,trim=0 20 0 0]{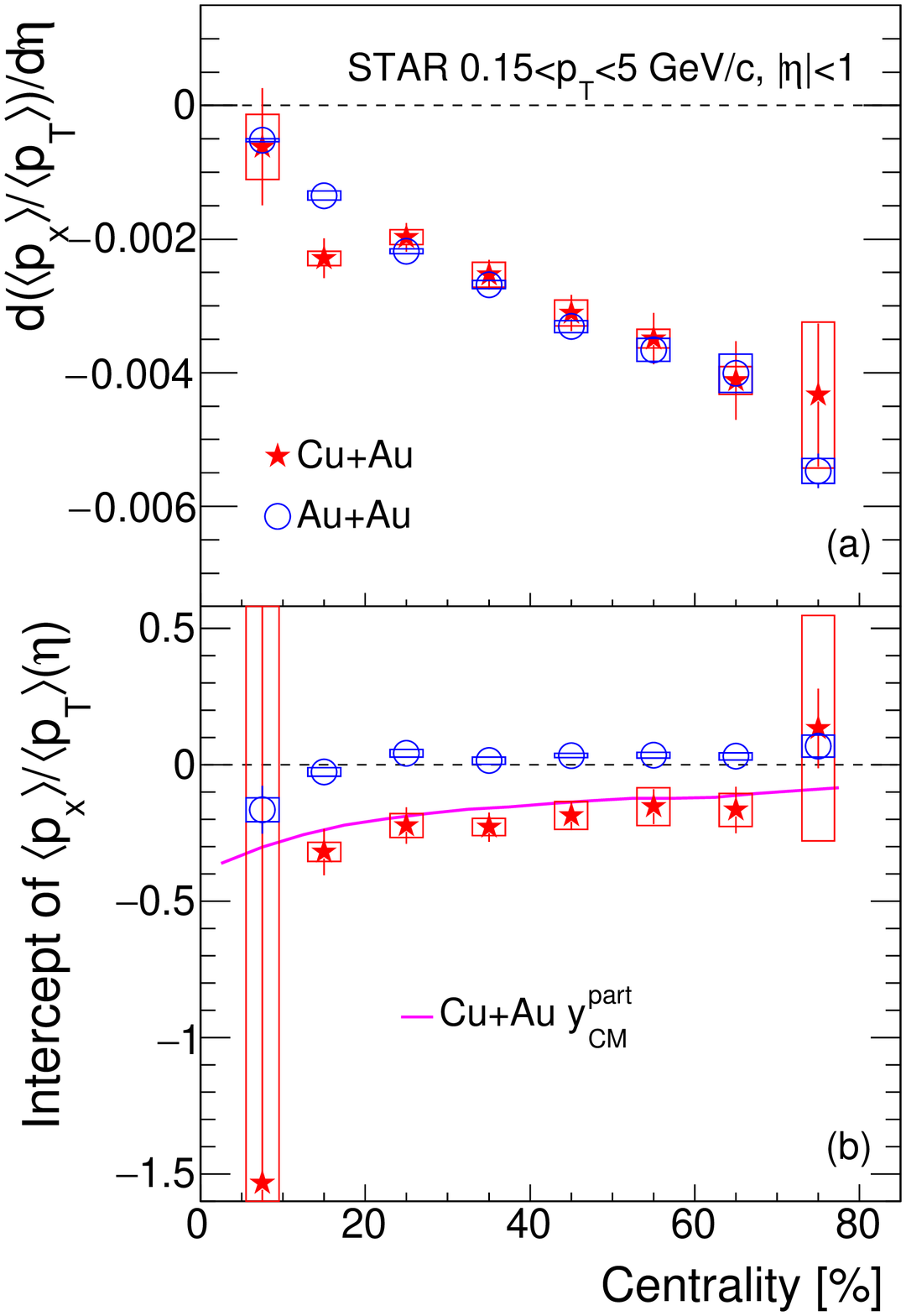}
\includegraphics[width=0.42\linewidth,trim=0 20 0 0]{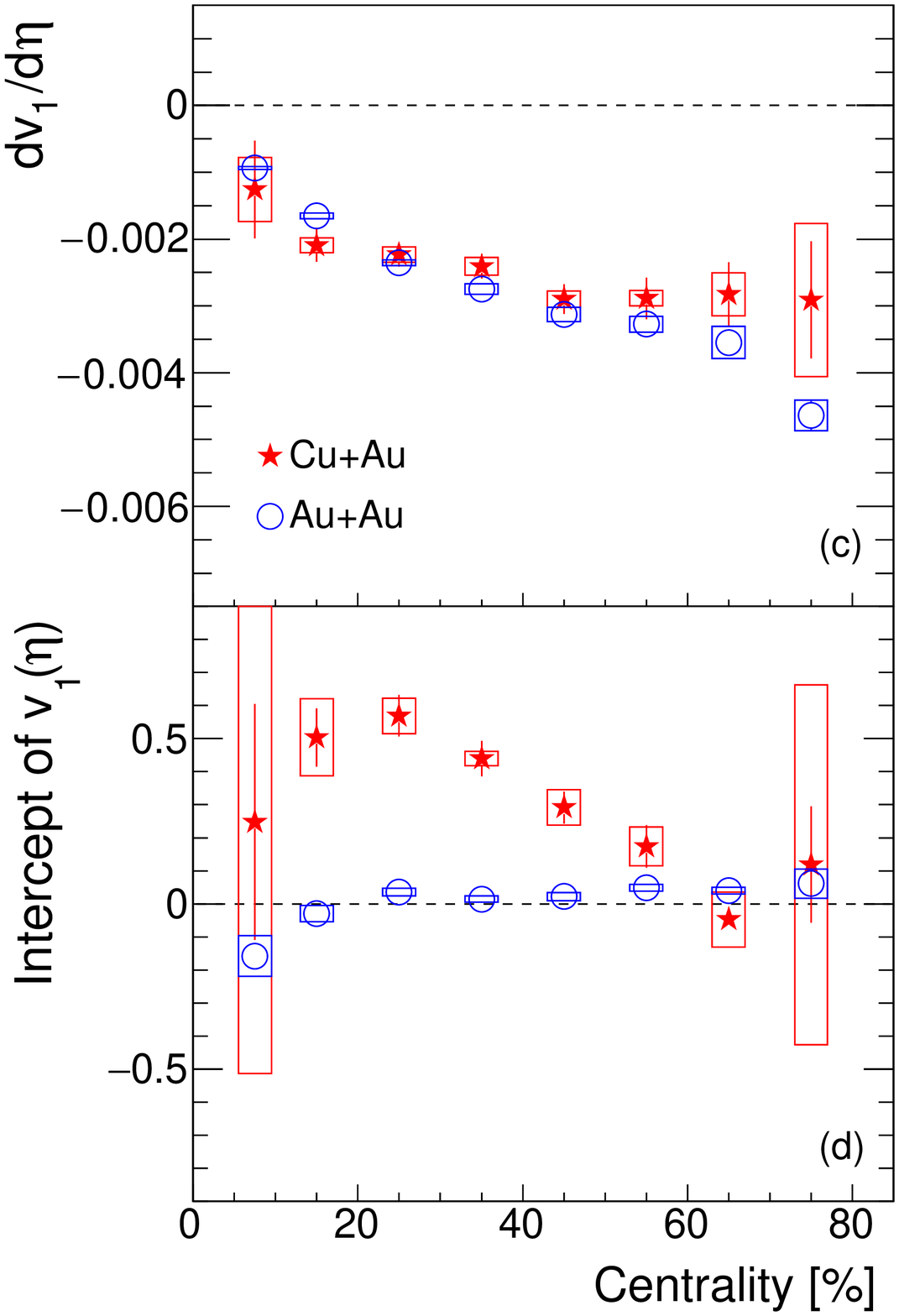}
\end{center}
\caption{(Color online) Slopes and intercepts of $\apx/\apt(\eta)$ and $v_1(\eta)$ 
  as a function of centrality in Cu+Au and Au+Au collisions at
  \sqsn = 200 GeV.  The solid line shows the center-of-mass rapidity
  in Cu+Au collisions calculated by Cu and Au participants in a
  Glauber model. Open boxes represent systematic uncertainties.}
\label{fig:slopes}
\end{figure*}
%

The results presented in Figs.~\ref{fig:v1eta_vsZDC} and \ref{fig:v1px_eta}, 
and in particular a positive intercept of $v_1({\eta})$ and 
negative intercept of $\apx$, are consistent with a picture of 
directed flow in Cu+Au collisions as a superposition of that from 
a ``tilted source" (shifted in rapidity to the system center-of-mass rapidity) 
and dipole flow due to non-zero average density gradients. 
Compared to the $v_1(\eta)$ dependence in symmetric collisions, 
the first mechanism shifts the function toward negative rapidities, 
and the second moves the entire function up (note that the Cu nucleus 
is defined as the projectile) as shown in Fig.~\ref{fig:tilt}(a,b).  
This picture receives further support from 
the study of the centrality dependence of the corresponding slopes and
intercepts presented in Fig.~\ref{fig:slopes}. Very similar slopes of
$v_1$ and \apx/\apt would be a natural consequence of a ``tilted source".
The intercepts of \apx follow very closely the shift in rapidity
center-of-mass of the system shown with the solid line in
Fig.~\ref{fig:slopes}(b), which was calculated by a Monte-Carlo
Glauber model based on the ratio of Au and Cu participant nucleons:
\begin{eqnarray}
y_{\rm CM} \approx \frac{1}{2}\ln(N_{\rm part}^{\rm Au}/N_{\rm part}^{\rm Cu}),
\label{eq:ycm}
\end{eqnarray}
where $N_{\rm part}^{\rm Au(Cu)}$ is the number of participants from
Au or Cu nuclei.  The centrality dependence of $v_1$ intercept (more
exactly, in this picture the difference in $v_1$ and \apx intercepts)
in Fig.~\ref{fig:slopes}(d) would be mostly determined by the
decorrelations between the dipole flow direction, $\Psi_{1,3}$, and
the reaction (spectator) planes.
 
The slopes of $v_{1}^{\rm odd (conv)}$ and $\langle p_x^{\rm conv}
\rangle/\mpt$ in Fig.~\ref{fig:v1px_eta} agree within 10\% both in
Au+Au and Cu+Au collisions. In Pb+Pb collisions at the LHC energy the
$v_1$ slope is almost a factor of two larger in magnitude than that of
$\langle p_x^{\rm conv} \rangle/\mpt$. This clearly indicates that
both mechanisms, ``tilted source'' (for which one would expect the
slope of $\langle p_x^{\rm conv} \rangle/\mpt$ to be about 50\% larger
than that of $v_{1}^{\rm odd (conv)}$, see Appendix), and initial
density asymmetries (for which $\langle p_x^{\rm conv} \rangle =0$),
play a significant role in the formation of the directed flow even in
symmetric collisions.  The relative contribution of the ``tilted
source" mechanism to the $v_1$ slope, $r$, can be expressed as (see
Appendix):
\begin{eqnarray}
r = \frac{\displaystyle \left( \frac{dv_1}{d\eta} \right)^{\rm tilt} }{\displaystyle \frac{dv_1}{d\eta} } 
  \approx \frac{2}{3} \frac{\displaystyle \frac{1}{\mpt} \frac{d\mpx}{d\eta} }{\displaystyle \frac{dv_1}{d\eta} },
\end{eqnarray}
where (~)$^{\rm tilt}$ denotes a contribution from the ``tilted
source".  The relative contribution $r$ is about 2/3 at the top RHIC
collision energies decreasing to about 1/3 at LHC energies. From the
centrality dependence of slopes shown in Fig.~\ref{fig:slopes} one can
conclude that the relative contribution of the ``tilted source"
mechanism is largest in peripheral collisions (where the $\langle
p_x^{\rm conv} \rangle/\mpt$ slope is approximately 1.5 times larger
than that of $v_{1}^{\rm odd (conv)}$) and smallest in central
collisions.  This dependence might be due to the stronger
decorrelation between spectator and dipole flow planes in peripheral
collisions.
%
\begin{figure}[ht]
\begin{center}
\includegraphics[width=\linewidth,trim=0 30 0 0]{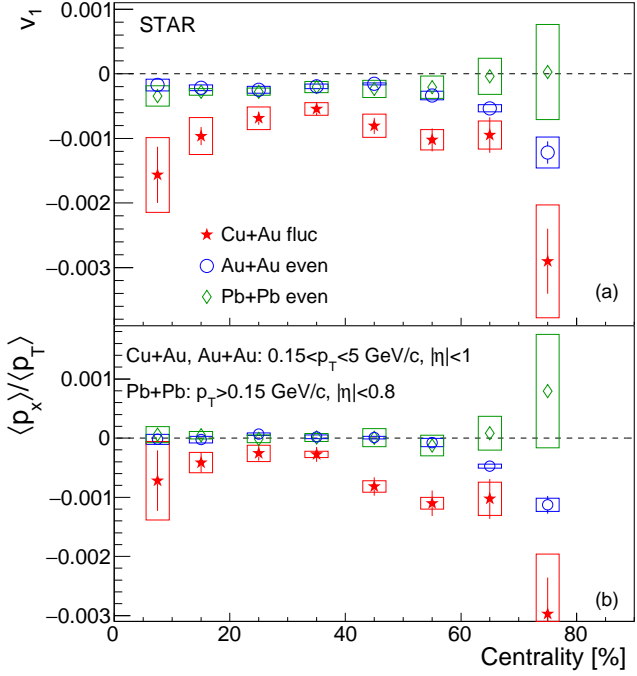}
\end{center}
\caption{(Color online) Centrality dependence of the even
  (fluctuation) components of $v_1$ and $\apx/\apt$ in Cu+Au and Au+Au
  collisions at \sqsn = 200 GeV and Pb+Pb collisions at \sqsn = 2.76
  TeV~\cite{aliceV1}. Open boxes represent systematic uncertainties.}
\label{fig:v1cent}
\end{figure}
%
\begin{figure*}[bth]
\begin{center}
\includegraphics[width=0.9\linewidth,trim=0 20 0 0]{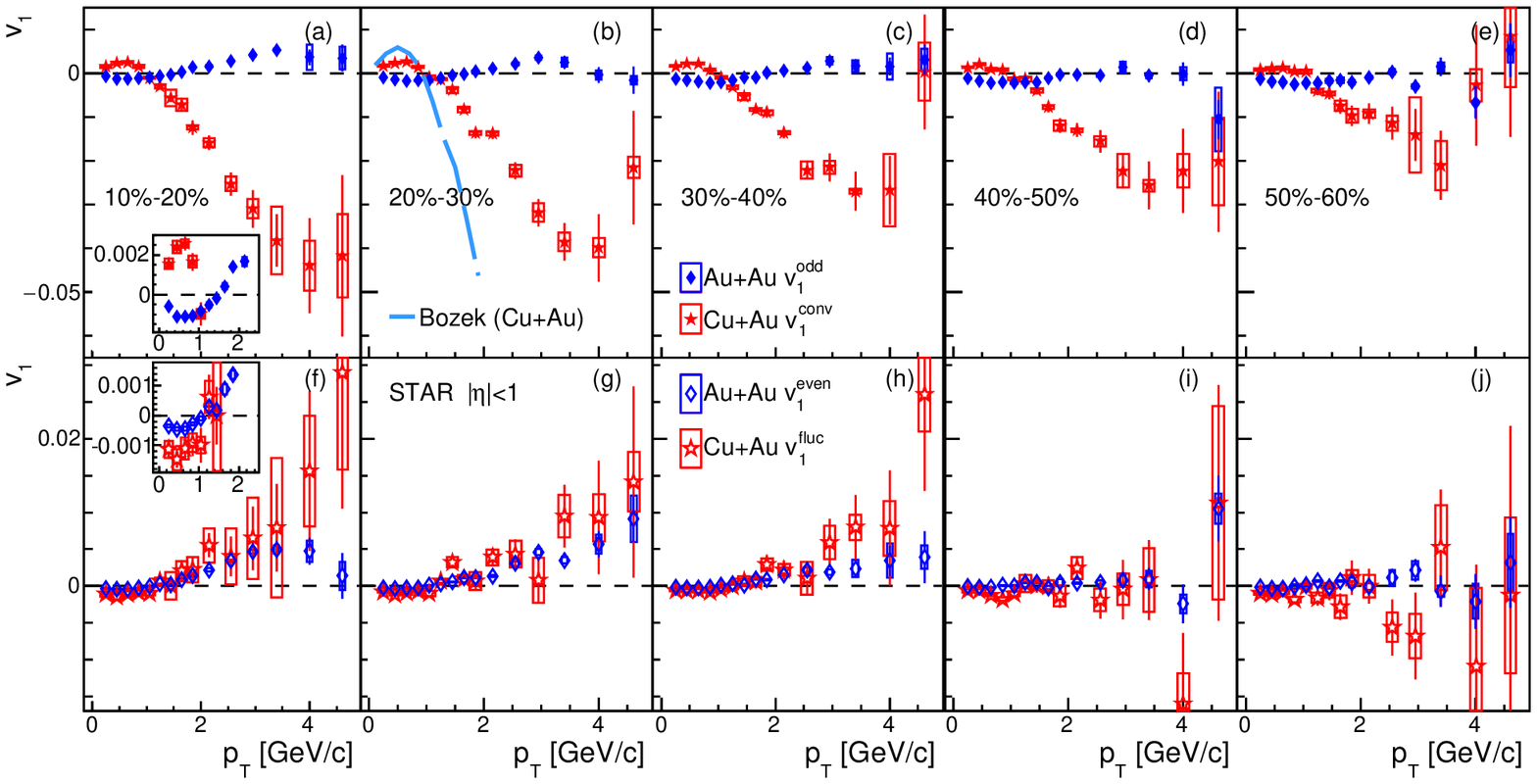}
\end{center}
\caption{(Color online) The conventional (a-e) and fluctuation (f-j)
  components of directed flow, $v_1^{\rm conv(odd)}$ and $v_1^{\rm
    fluc(even)}$, of charged particles as a function of \pt for
  different collision centralities in Cu+Au and Au+Au collisions. Open
  boxes represent systematic uncertainties.  The broken line in panel
  (b) shows the viscous hydrodynamic calculation for Cu+Au
  collisions~\cite{bozek}.}
\label{fig:v1pt_cent}
\end{figure*}
Figure~\ref{fig:v1cent} shows the even (fluctuation) components of
$v_1$ and $\apx$ as a function of centrality. The $v_1^{\rm even}$ for
Au+Au has a weak centrality dependence and is consistent with
$v_1^{\rm even}$ for Pb+Pb except in most peripheral collisions.
Furthermore, $p_x^{\rm even}$ in both Au+Au and Pb+Pb are consistent
with zero.  This may indicate that the dipole-like fluctuation in the
initial state has little dependence on the system size and collision
energy.  $v_1^{\rm fluc}$ and $\apx^{\rm fluc}$ for Cu+Au has a larger
magnitude than in symmetric collisions over the entire centrality
range; it is smallest in the 30\%-40\% centrality bin.

%
\begin{figure}[htb]
\begin{center}
\includegraphics[width=\linewidth,trim=0 25 0 0]{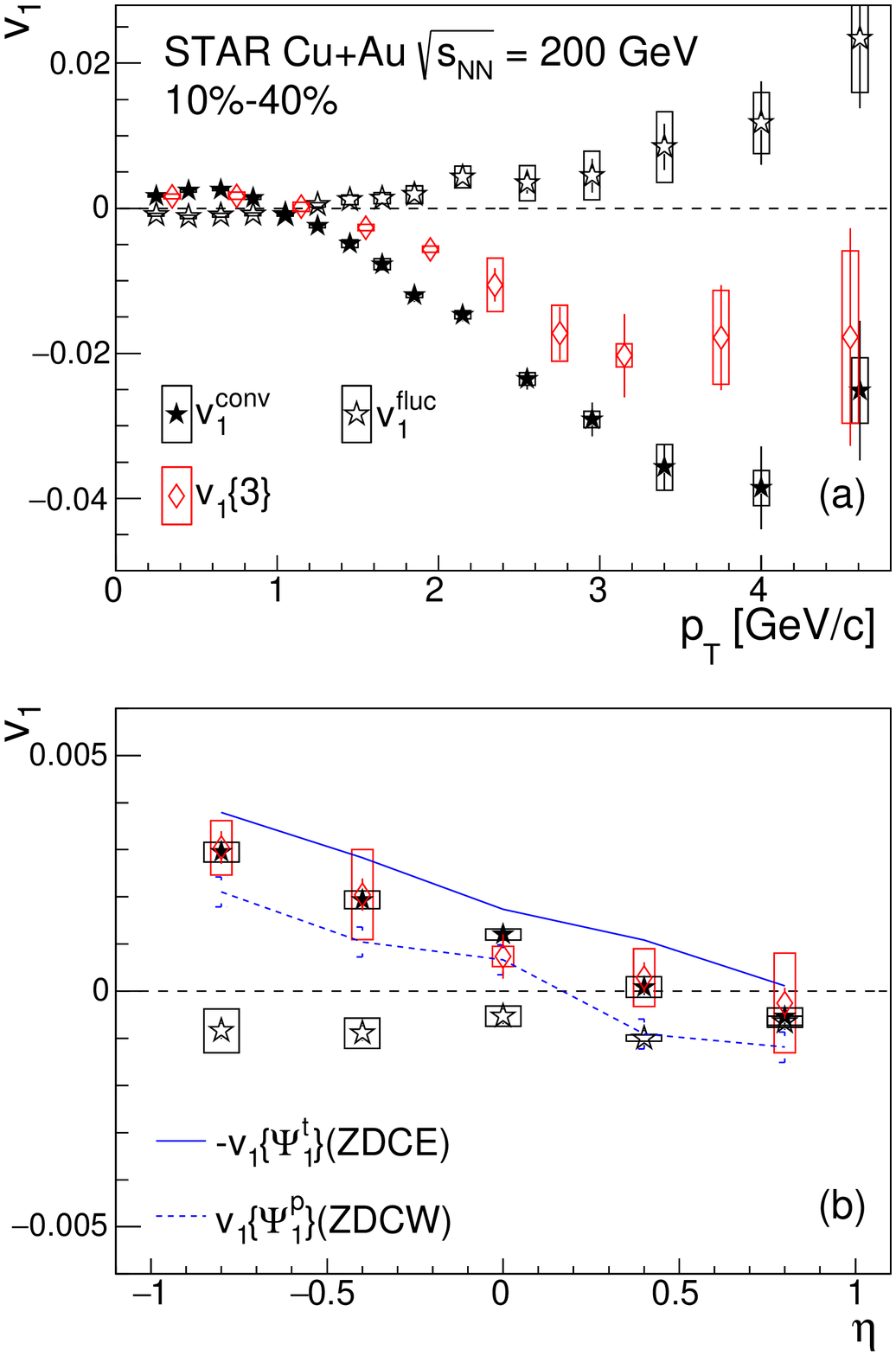}
\end{center}
\caption{(Color online) Directed flow of charged particles as a
  function of \pt (a) and $\eta$ (b) in the 10\%-40\% centrality bin
  measured with the ZDC-SMD event planes and three-point correlator in
  Cu+Au collisions. The \pt dependence was measured in $|\eta|<1$ and
  the $\eta$ dependence was integrated over $0.15<\pt<5$ GeV/$c$. Open
  boxes represent systematic uncertainties.}
\label{fig:v1_3p}
\end{figure}
The reference angle of dipole flow can be represented by $\Psi_{1,3}$,
but $v_1^{\rm even}$ ($v_1^{\rm fluc}$) are the projections of dipole
flow onto the spectator planes. Therefore, the measured even (or
fluctuation) components of $v_1$ should be decreased by a factor
$\langle\cos(\Psi_{1,3}-\Psi_{\rm SP})\rangle$.  Such a ``resolution''
effect may also lead to larger $v_1^{\rm even}$ and non-zero $\langle
p_x^{\rm even}\rangle$ in Cu+Au collisions due to the difference in
correlation of the Cu and Au spectator planes to $\Psi_{1,3}$.

The \pt dependence of $v_1^{\rm conv}$ and $v_1^{\rm fluc}$ in Cu+Au
collisions was studied for different collision centralities, as shown
in Fig.~\ref{fig:v1pt_cent}. The $v_1^{\rm conv}$ exhibits a sign
change around \pt = 1 GeV/$c$ and its magnitude at both low and high
\pt becomes smaller for peripheral collisions.  Such centrality
dependence in Cu+Au $v_1^{\rm conv}$ can be due to a change in the
correlation between the angle of the initial density asymmetry and the
direction of spectator deflection. The correlation becomes largest at
an impact parameter of 5~fm (which corresponds approximately to
10\%-20\% centrality) and decreases in more peripheral collisions as
discussed in Ref.~\cite{Voloshin:2016ppr}.  Similar \pt and centrality
dependencies were observed in $v_1^{\rm fluc}$ although there is a
difference in sign between $v_1^{\rm conv}$ and $v_1^{\rm fluc}$.  An
event-by-event viscous hydrodynamic model calculation is also compared
to the $v_1^{\rm conv}$ for the 20\%-30\% centrality bin in Cu+Au
collisions. As seen in Fig.~\ref{fig:v1pt_cent}, the model
qualitatively follows the shape of the measurement but overpredicts
the data in its magnitude for the entire $p_T$ region.

The odd and even components of directed flow, $v_1^{\rm odd}$ and
$v_1^{\rm even}$, in Au+Au collisions are also compared in the same
centrality windows, where $v_1^{\rm odd}$ was measured by flipping the
sign for particles with the negative rapidity.  The signals of both
$v_1^{\rm odd}$ and $v_1^{\rm even}$ in Au+Au are smaller than
directed flow in Cu+Au but, at least in central collisions, they still
show the sign change in the \pt dependence.

The $v_1$ with the three-point correlator, $v_1\{3\}$, was measured in
Cu+Au collisions for the 10\%-40\% centrality bin as shown in
Fig.~\ref{fig:v1_3p}, where it is compared to $v_1^{\rm conv}$ and
$v_1^{\rm fluc}$ from the event plane method using spectator
planes. Note that $v_1\{3\}$ does not use spectator information.  The
$v_1\{3\}$ is consistent with $v_1^{\rm conv}$ for $\pt<1$ GeV/$c$
within the systematic uncertainties but becomes greater than $v_1^{\rm
  conv}$ for $1<\pt<4$ GeV/$c$.
The $v_{1}\{3\}$ includes both conventional and fluctuation components
of $v_1$.  The conventional component in $v_1\{3\}$ should be the same
as measured by the event plane method but the fluctuation component
might be different due to different correlations of the spectator
planes and participant plane (from the BBC subevent) with
$\Psi_{1,3}$.

\subsection{Directed flow of identified hadrons\label{sec:pidv1}}

\begin{figure}[!htb]
\begin{center}
\includegraphics[width=\linewidth,trim=0 20 0 0]{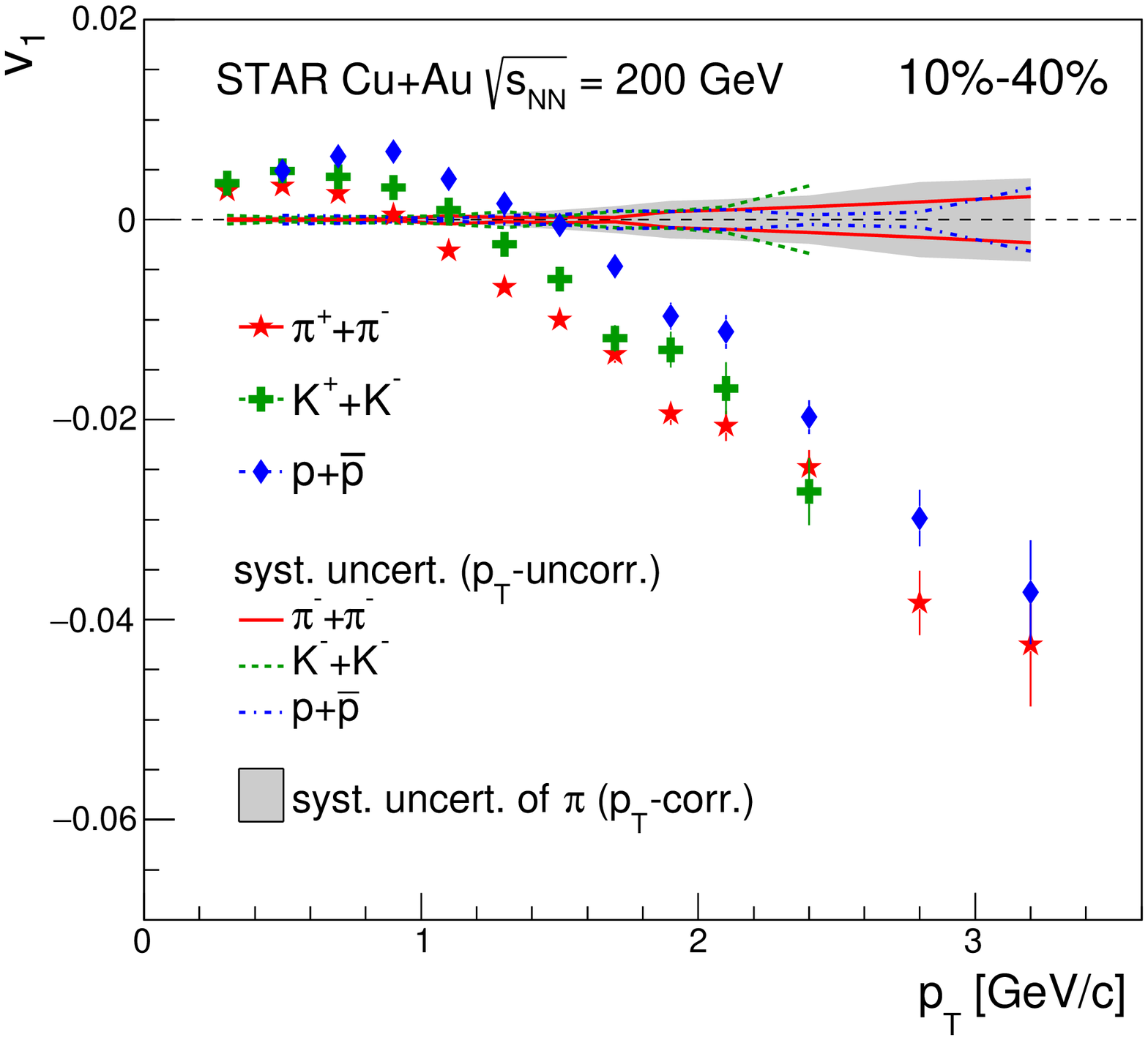}
\end{center}
\caption{(Color online) Directed flow of $\pi^{+}+\pi^{-}$, $K^{+}+K^{-}$,
  and $p+\bar{p}$ as a function of \pt for $|\eta|<1$ in the 10\%-40\% centrality
  bin. The \pt-uncorrelated systematic uncertainties are shown with lines
  around $v_1=0$ for each particle species.  \pt-correlated systematic
  uncertainty is shown only for pions with a shaded band.}
\label{fig:pidv1}
\end{figure}

Anisotropic flow of charged pions, kaons, and (anti)protons was
measured based on the particle identification with the TPC and TOF, as
explained in Sec.~\ref{sec:track}. Figure~\ref{fig:pidv1} presents
directed flow of $\pi^{+}+\pi^{-}$, $K^{+}+K^{-}$, and $p+\bar{p}$
measured with respect to the target (Au) spectator plane
($v_1=-v_1\{\Psi_{\rm SP}^t\}$) in the 10\%-40\% centrality bin. For
$\pt<2$~GeV/$c$, there is a clear particle type dependence, likely
reflecting the effect of particle mass in interplay of the radial and
directed flow~\cite{pidv1_AGS,Voloshin:1996nv}.  In $\pt>2$~GeV/$c$
region, there is no clear particle type dependence due to the large
uncertainties.  Measurement of identified particle $v_1$ with the
projectile (Cu) spectator plane is difficult due to small statistics
of identified particles and poor event plane resolution; therefore we
do not decompose the $v_1$ into the conventional and fluctuation
components. The presented $v_1$ of $\pi^{+}+\pi^{-}$, $K^{+}+K^{-}$,
and $p+\bar{p}$ includes both components.  The observed mass
dependence in the $v_1$ of identified particles is consistent with
results from the PHENIX Collaboration~\cite{cuauphenix}.
%

\subsection{Charge dependence of directed flow}
\begin{figure}[!thb]
\begin{center}
\includegraphics[width=\linewidth,trim=0 20 0 0]{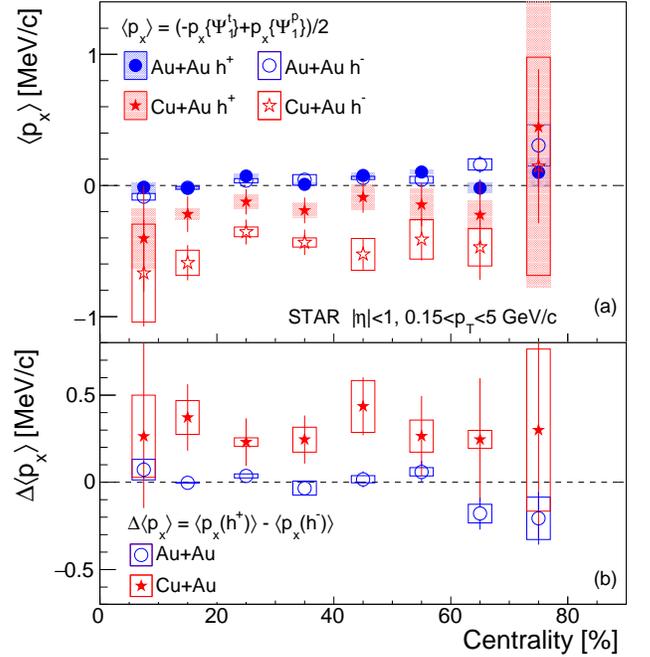}
\end{center}
\caption{(Color online) Positively and negatively charged particles
  $\apx$ and the difference $\Delta\apx$ as a function of centrality
  in Au+Au and Cu+Au collisions. Open and shaded boxes show systematic
  uncertainties.  }
\label{fig:chpx}
\end{figure}

In our previous study~\cite{cuauv1_star}, a finite difference in
directed flow between positively and negatively charged particles was
observed in asymmetric Cu+Au collisions.  These results can be
understood as an effect of the electric field due to the asymmetry in
the electric charge of the Au and Cu nuclei.  Similarly, one would
expect a difference in $\apx$ between positive and negative particles.
Figure~\ref{fig:chpx} shows the centrality dependence of
charge-dependent $\apx$ and the difference $\Delta\apx$ between
positive and negative particles in Au+Au and Cu+Au collisions.  The
difference is consistent with zero for Au+Au collisions, but a finite
difference is observed in Cu+Au collisions
($\Delta\apx\sim$0.3~MeV/$c$).  The direction of the electric field is
expected to be strongly correlated to the direction of the Cu
(projectile) spectator deflection, which should lead to a positive
$\apx$ by the convention used in this analysis.  The results are
consistent with these expectations.

%
The magnitude of the momentum shift can be roughly estimated based on
the equation of motion, i.e. $\Delta p_x = e|\vec{E}|/m_{\pi}^2\times
m_{\pi}^{2}\times\Delta t$ where $\vec{E}$ denotes the electric field,
$m_{\pi}$ is a pion mass, and $\Delta t$ is the lifetime of the
electric field.  If one takes $e|\vec{E}|/m_{\pi}^2\sim0.9$ and
$\Delta t \sim 0.1$~fm/$c$~\cite{voronyuk}, assuming that the time
dependence of the electric field approximates a step function, the
resulting $\Delta p_x$ is $\sim$9 MeV/$c$ which is $\sim$30 times
larger than the observed $\Delta\apx$. The charge-dependence of
$\Delta \apx$ is determined by the number of charges, i.e. the number
of quarks and antiquarks, at the time when the initial electric field
is strong after the collisions.  Therefore a difference in $\Delta
\apx$ between the data and our estimate might indicate a smaller
number of quarks and antiquarks at early times ($t<0.1$~fm/$c$)
compared to the number of quarks in the final state, as discussed in
Ref.~\cite{cuauv1_star}. The lifetime of the electric field depends on
the model and could be longer if the medium has a larger conductivity.
Also note that the observed $\Delta \apx$ might be smeared by the
fluctuations between the direction of the electric field and the
spectator plane, and by hydrodynamic evolution and hadron rescattering
at later stages of the collisions.

For a mere detailed view of the quark-antiquark production
dynamics, as well as to understand the role of baryon stopping in
the development of directed flow at midrapidity we also extended our
measurements to identified particles.  In the so-called ``two-wave''
scenario of quark production~\cite{Pratt:2013xca},  
the number of s quarks approximately remains the same during the
system evolution while the number of u and d quarks sharply
increases at the hadronization time.  In this case, one might
expect a relatively larger effect of the initial electric field for
s quarks than that for u and d quarks.  Therefore the measurement
of charge-dependent $v_1$ for pions and kaons might serve as a test
of such a quark production scenario. The difference in number of
protons and neutrons in the colliding nuclei in combination with the
baryon stopping might also contribute to the charge dependence of
directed flow. In this case one can expect a significantly larger
effect measuring the flow of baryons itself. For that we measure
the charge dependence of directed flow of protons and antiprotons.

\begin{figure}[!htb]
\begin{center}
\includegraphics[width=\linewidth,trim=0 0 0 0]{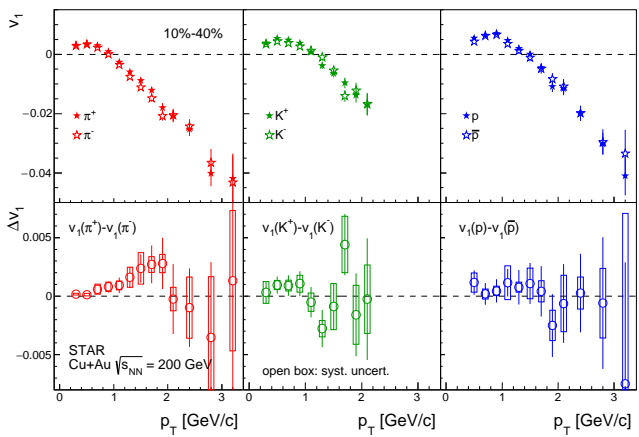}
\end{center}
\caption{(Color online) Directed flow of $\pi^{+}$, $\pi^{-}$,
  $K^{+}$, $K^{-}$, $p$, and $\bar{p}$ measured in $|\eta|<1$ as a function of \pt for the
  10\%-40\% centrality bin in Cu+Au collisions (top panels), where
  only the statistical uncertainties are shown.  The differences in
  the directed flow between positively and negatively charged
  particles are shown in bottom panels, where the open boxes show the
  systematic uncertainties.  }
\label{fig:pidDv1}
\end{figure}

Top panels in Fig.~\ref{fig:pidDv1} show \pt dependence of
$v_1$ separately for $\pi^{+}$ and $\pi^{-}$, $K^{+}$ and $K^{-}$,
and $p$ and $\bar{p}$ for 10\%-40\% centrality in Cu+Au collisions.
Bottom panels show difference in $v_1$, $\Delta v_1$, between
positively and negatively charged particles for each species.
Similarly as observed for charged hadrons~\cite{cuauv1_star} and in
agreement with results presented in Fig.~\ref{fig:chpx}, $v_1$ of
$\pi^{+}$ is larger than that of $\pi^{-}$ in the $\pt<2$ GeV/$c$
region, which is consistent with the expectation from the initial
electric field effect.  For charged kaons and (anti)protons, no
significant difference are observed within the current experimental
precision. 

%

\section{Elliptic and higher harmonic flow\label{sec:vn}}
\subsection{Unidentified charged particles}

\begin{figure*}[th]
\begin{center}
\includegraphics[width=\linewidth,trim=0 20 0 0]{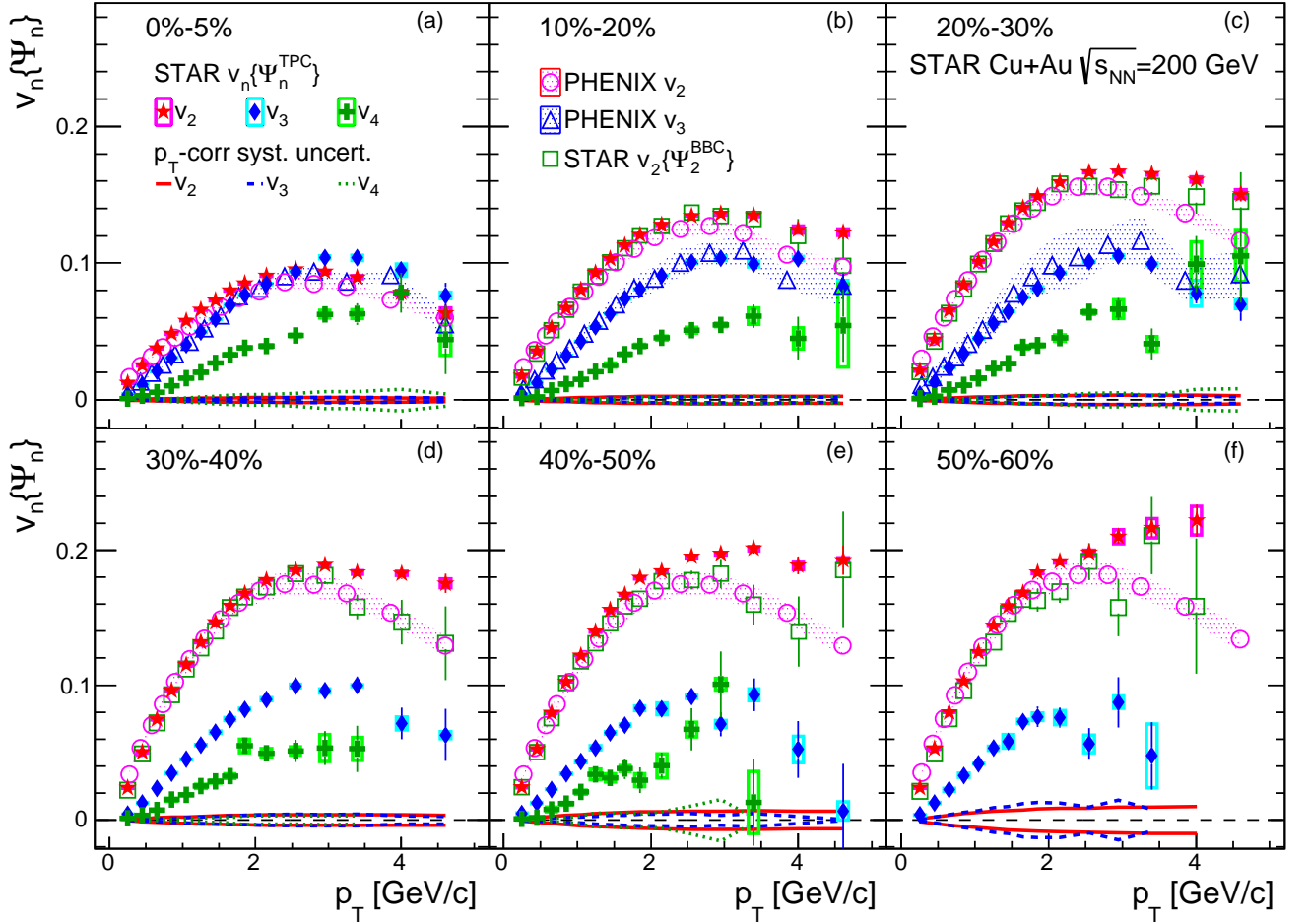}
\end{center}
\caption{(Color online) Higher harmonic flow coefficients
  $v_n\{\Psi_n\}$ of charged particles in Cu+Au collisions as a
  function of \pt for six centrality bins.  Colored boxes around the
  data points show \pt-uncorrelated systematic uncertainties and solid
  thin lines around $v_n=0$ show \pt-correlated systematic
  uncertainties.  Results from the PHENIX experiment~\cite{cuauphenix}
  are compared. Only statistical uncertainty is shown for $v_{2}\{\Psi_2^{\rm BBC}\}$.  }
\label{fig:vnpt}
\end{figure*}

Higher harmonic anisotropic flow coefficients, $v_n$, of charged
particles were measured with TPC $\eta$ subevents as a function of \pt
up to $n=4$.  Results for six centrality bins (0\%-5\%, 10\%-20\%,
20\%-30\%, 30\%-40\%, 40\%-50\%, and 50\%-60\%) are shown in
Fig.~\ref{fig:vnpt}.  Results for $v_2$ and $v_3$ from the PHENIX
experiment~\cite{cuauphenix}, shown for comparison, agree well with
our results within uncertainties.  The small difference in $v_2$ for
$\pt>2$~GeV/$c$ can be explained by a different contribution from
non-flow correlations -- PHENIX measured $v_2$ with a larger $\eta$
gap ($\Delta\eta>$2.65) between the particles of interest and those
used for the event plane determination, while our TPC $\eta$ subevents
have $\Delta\eta>$0.4. To confirm that explanation, we also calculated
$v_2$ with respect to the BBC event plane, which ensures
$\Delta\eta>2.3$. Those results, while having larger statistical
uncertainties, are consistent with the PHENIX measurements.
%

As with Au+Au collisions~\cite{v2star,v3star,vnphenix}, 
the elliptic flow, $v_2$, in Cu+Au collisions depends strongly on centrality,
increasing significantly toward more peripheral collisions. The $v_3$ and
$v_4$ have weak centrality dependencies.  In the most central
collisions, the magnitude of $v_3$ is comparable to, or even greater
than, $v_2$ for $\pt>2$ GeV/$c$. A similar trend has been observed at
the LHC~\cite{vn_alice}.

\begin{figure}[htb]
\begin{center}
\includegraphics[width=0.9\linewidth,trim=0 20 0 0]{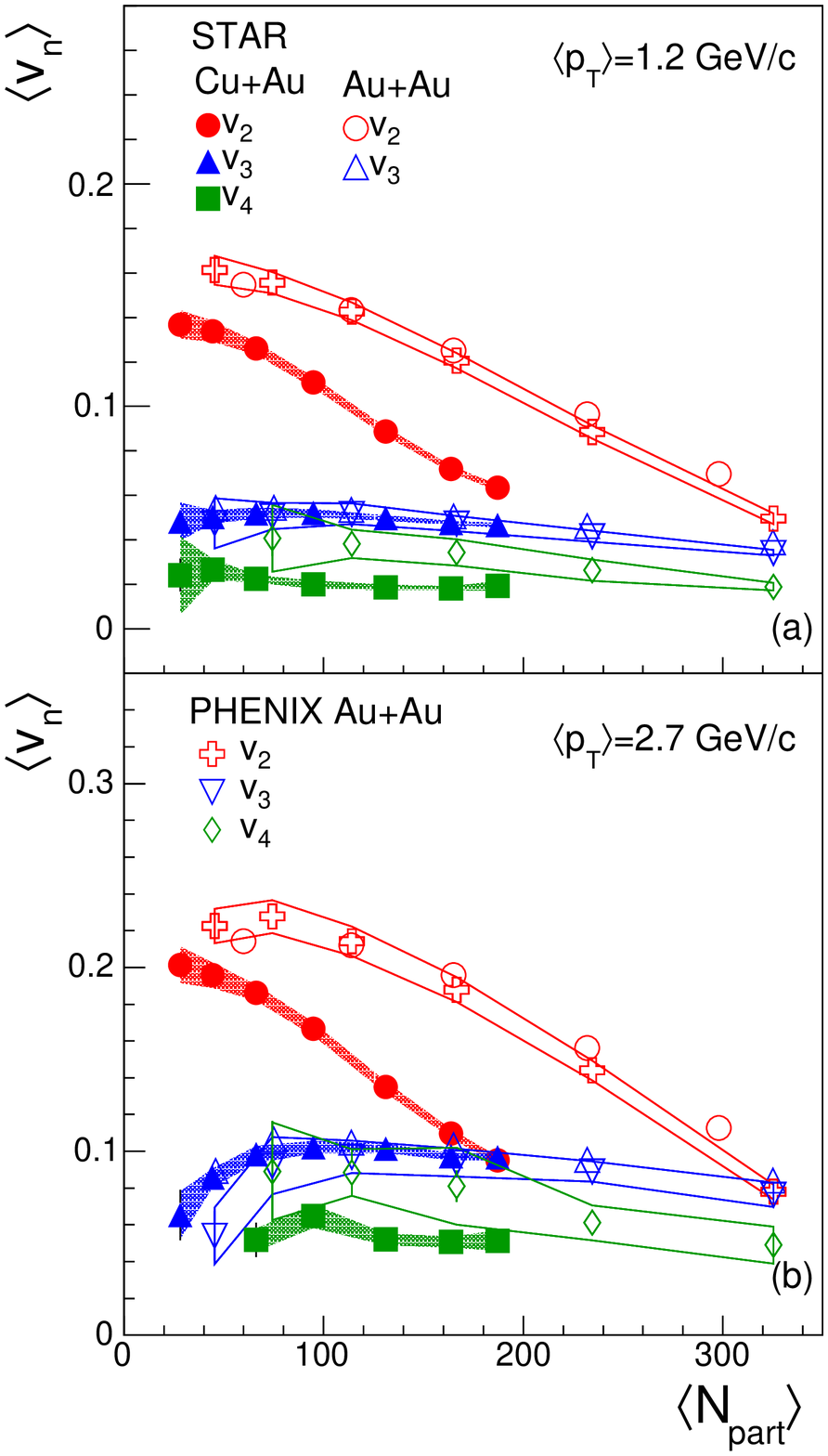}
\end{center}
\caption{(Color online) Higher harmonic flow coefficients $v_n$ of charged
  particles for two selected \pt bins as a function of the number of
  participants calculated with a Monte-Carlo Glauber simulation for
  Cu+Au and Au+Au collisions, comparing with results in Au+Au from the PHENIX experiment~\cite{vnphenix}.  
  Open and shaded bands represent systematic uncertainties.  }
\label{fig:vn_npart}
\end{figure}

To make a comparison with Au+Au collisions, the Cu+Au results are
plotted as a function of the number of participants for two different
\pt bins in Fig.~\ref{fig:vn_npart}.  Results for Au+Au collisions
were taken from the previous studies by STAR~\cite{v2star,v3star} and
PHENIX~\cite{vnphenix}. The elliptic flow, $v_2$, has a strong centrality
dependence in both systems due to the variation of the initial
eccentricity, while $v_3$ and $v_4$ have much weaker centrality
dependence reflecting their mostly fluctuation origin.  The triangular flow, $v_3$, 
as a function of the number of participants in Cu+Au falls on the same
curve as in Au+Au.  This suggests that $v_3$ (determined by the
initial triangularity) is dominated by fluctuations, which are
directly related to the number of participants. The $v_4$ in Au+Au is
slightly larger than in Cu+Au.  These relations between $v_n$ in the
two systems can be qualitatively explained by the initial spatial
anisotropy, $\varepsilon_n$~\cite{en_glasma}. A larger $v_4$ in Au+Au
collisions compared to that in Cu+Au may be due to a larger $v_2$ and
$v_2$-$v_4$ nonlinear coupling that cannot be fully accounted for by
the $\varepsilon_2$-$\varepsilon_4$ correlation~\cite{vn_atlas}.

%
%
\begin{figure}[thb]
\begin{center}\vspace{0.3cm}
\includegraphics[width=\linewidth,trim=0 20 0 0]{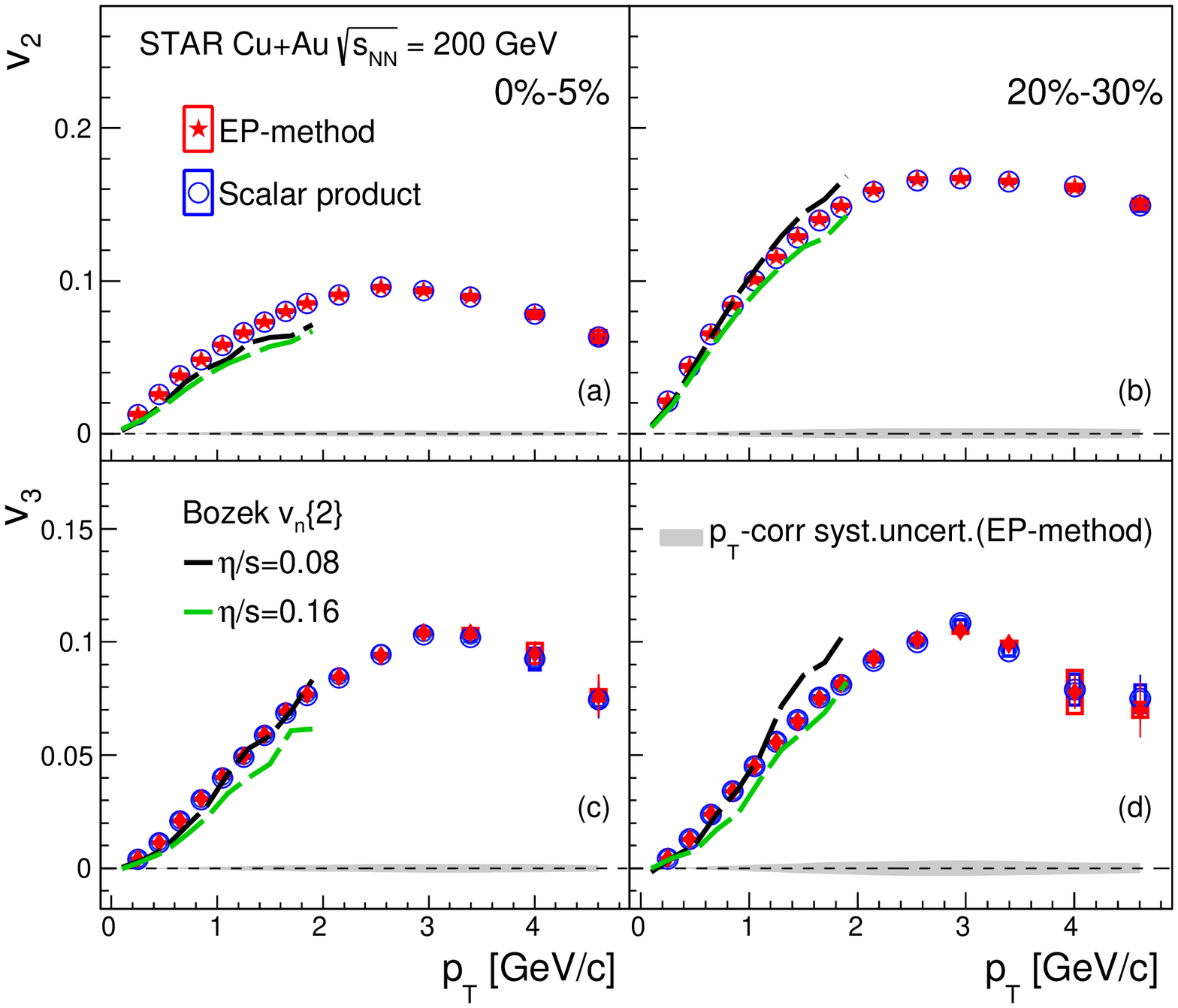}
\end{center}
\caption{(Color online) The second and third harmonic flow coefficients of charged particles as a
  function of \pt measured with the event plane (EP) method and scalar product method, comparing
  to the viscous hydrodynamic calculations~\cite{bozek}.
  Panels (a,c) are for 0\%--5\% centrality, panels (b,d) for 20\%-30\% centrality.}
\label{fig:vn_hydro}
\end{figure}

Hydrodynamic models have successfully described the azimuthal
anisotropy measured in symmetric collisions.  The comparison of the
data to model calculations provided valuable constraints on the shear
viscosity over entropy density $\eta/s$~\cite{vnphenix,schenke}.
Further constraints can be obtained from a similar comparison for
asymmetric collisions.  Figure~\ref{fig:vn_hydro} compares $v_2$ and
$v_3$ in Cu+Au collisions to the viscous hydrodynamic
calculations~\cite{bozek}.  The model employs the Glauber (participant
nucleons) initial density distribution and applies the event-by-event
viscous hydrodynamic model with $\eta/s$ = 0.08 or 0.16.  Both $v_2$
and $v_3$ are reasonably well described by the model at $\pt<2$ GeV/$c$. The calculation
with $\eta/s$ = 0.08 seems to work better in the 0\%-5\% centrality
bin, while the 20\%-30\% centrality results might need a larger
$\eta/s$.  In the same figure we also compare $v_2$ and $v_3$ measured
with the scalar product method to the corresponding measurements
obtained with the event plane method.  Both methods use TPC $\eta$
subevents. The results are in a very good agreement with each other.

%
\begin{figure*}[bht]
\begin{center}
\includegraphics[width=0.95\linewidth,trim=0 20 0 0]{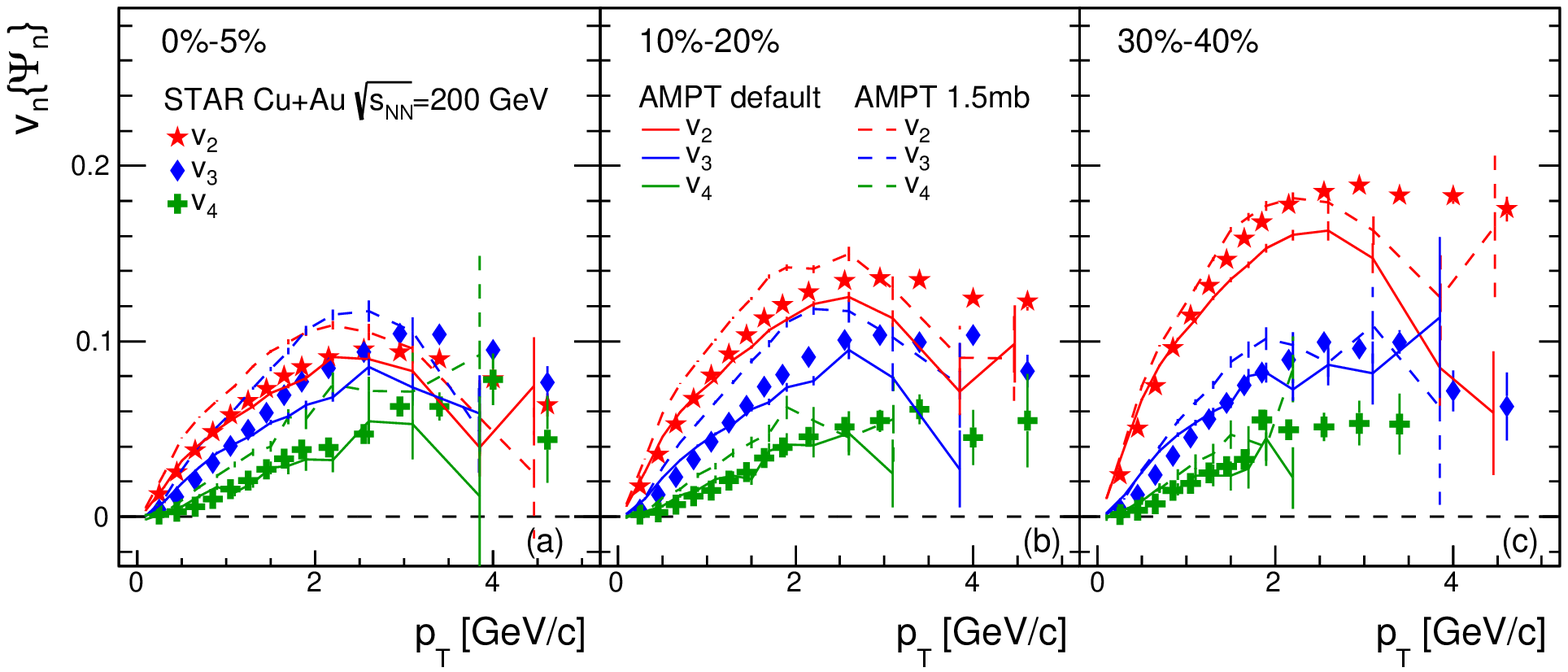}
\end{center}
\caption{(Color online) Higher harmonic flow coefficients $v_n$ of charged
  particles as a function of \pt comparing to the AMPT model~\cite{Lin:2004en}, 
  where solid lines are for default AMPT setup and dashed 
  lines are for the string melting version with $\sigma_{\rm parton}=1.5$ mb.}
\label{fig:vn_ampt}
\end{figure*}

Figure~\ref{fig:vn_ampt} compares our results to a multi-phase
transport (AMPT) model~\cite{Lin:2004en} (v1.26t5 for the default version and v2.26t5
for the string melting version).  The initial conditions in this model
are determined by the Heavy Ion Jet Interaction Generator (HIJING)~\cite{Wang:1991hta}
which is based on the Glauber model and creates minijet partons and
excited strings. In the AMPT default version, the strings are
converted into hadrons via string fragmentation, while in the string
melting version the strings are first converted to partons
(constituent quarks) and the created partons are converted to hadrons
via a coalescence process after the subsequent parton scatterings.

The event plane and centrality in the model calculations were
determined in the same way as in the real data analysis.  Flow
measurements were also performed in the same way.
Figure~\ref{fig:vn_ampt} shows $v_n$ for the 0\%-5\%, 10\%-20\%, and 30\%-40\%
centrality bins compared to the AMPT model in the default and string
melting versions. The parton cross section in the string melting
version was set to $\sigma_{\rm parton}=1.5$~mb~\cite{Xu:2011fe,Sun:2015pta}.
The AMPT calculations with the default version and the string melting
version with $\sigma_{\rm parton}=1.5$~mb qualitatively describe the data of
$v_2$, $v_3$, and $v_4$ for $\pt<3$ GeV/$c$.  The data is between the
default and string melting with $\sigma_{\rm parton}=1.5$~mb results, similar
to the observation in Ref.~\cite{Sun:2015pta,bes_pidv2}.

\subsection{Flow of identified hadrons and NCQ scaling\label{sec:pidvn}}

\begin{figure*}[htb]
\begin{center}
\includegraphics[width=0.95\linewidth,trim=0 20 0 0]{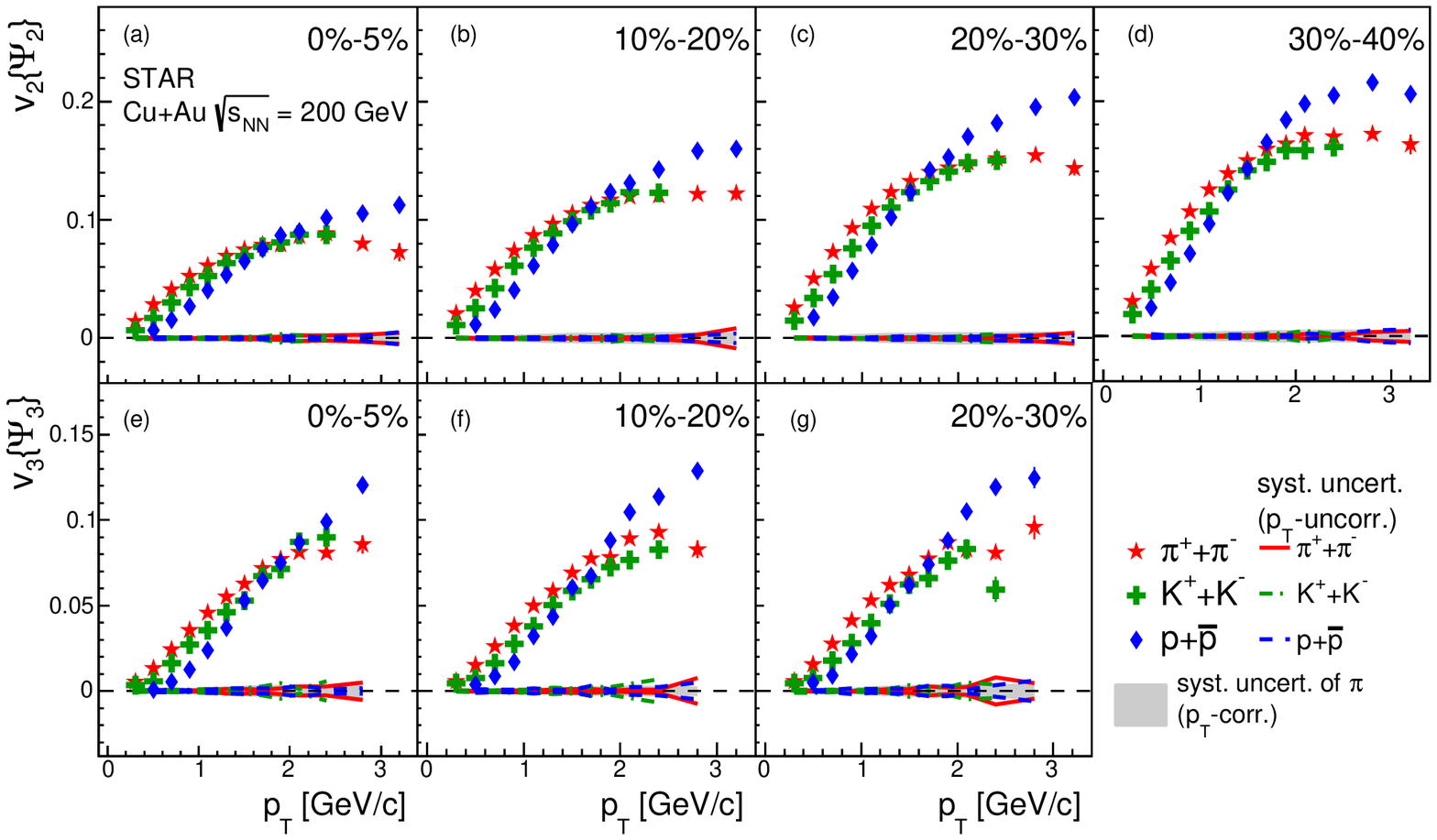}
\end{center}
\caption{(Color online) The second and third harmonic flow
  coefficients of $\pi^{+}+\pi^{-}$, $K^{+}+K^{-}$, and $p+\bar{p}$ as
  a function of \pt for four centrality bins.  Solid lines represent
  \pt-uncorrelated systematic uncertainties for each species.  Shaded
  bands represent \pt-correlated systematic uncertainties for pions.}
\label{fig:pidvn_cent}
\end{figure*}

\begin{figure*}[htb]
\begin{center}
\includegraphics[width=0.95\linewidth,trim=0 20 0 0]{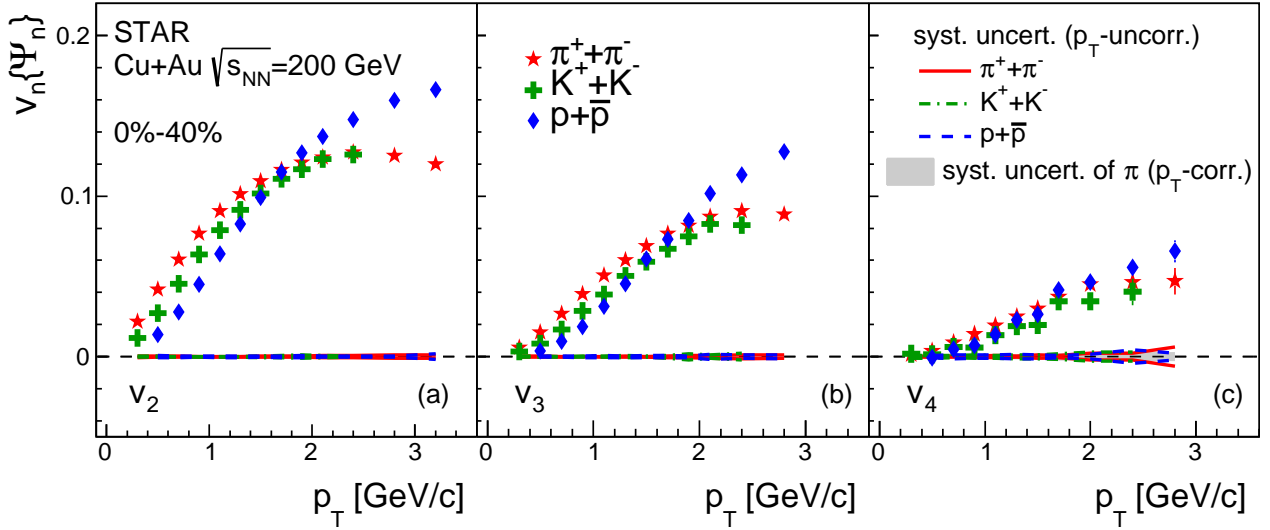}
\end{center}
\caption{(Color online) Higher harmonic flow coefficients $v_n$ of $\pi^{+}+\pi^{-}$,
  $K^{+}+K^{-}$, and $p+\bar{p}$ as a function of \pt in the 0\%-40\%
  centrality bin.
  Solid lines represent \pt-uncorrelated systematic uncertainties for each species.
Shaded bands represent \pt-correlated systematic uncertainties for pions.}
\label{fig:pidvn}
\end{figure*}

Anisotropic flow of charged pions, kaons, and (anti)protons was also
measured for higher harmonics ($n=$~2--4).  Figure~\ref{fig:pidvn_cent}
presents $v_2$ and $v_3$ of $\pi^{+}+\pi^{-}$, $K^{+}+K^{-}$, and
$p+\bar{p}$ for different centralities. A particle mass dependence is
clearly seen at low transverse momenta ($\pt<1.6$ GeV/$c$) similar to
that seen in $v_1$ in Fig.~\ref{fig:pidv1}.  In the \pt range
$1.6<\pt<3.2$ GeV/$c$, the splitting between baryons and mesons is
observed in $v_2$ and $v_3$.  Results for a wide centrality bin
(0\%-40\%) are shown in Fig.~\ref{fig:pidvn}, along with results for
$v_4$ that show similar trends to $v_2$ and $v_3$.

\begin{figure*}[htb]
\begin{center}
\includegraphics[width=0.95\linewidth,trim=0 20 0 0]{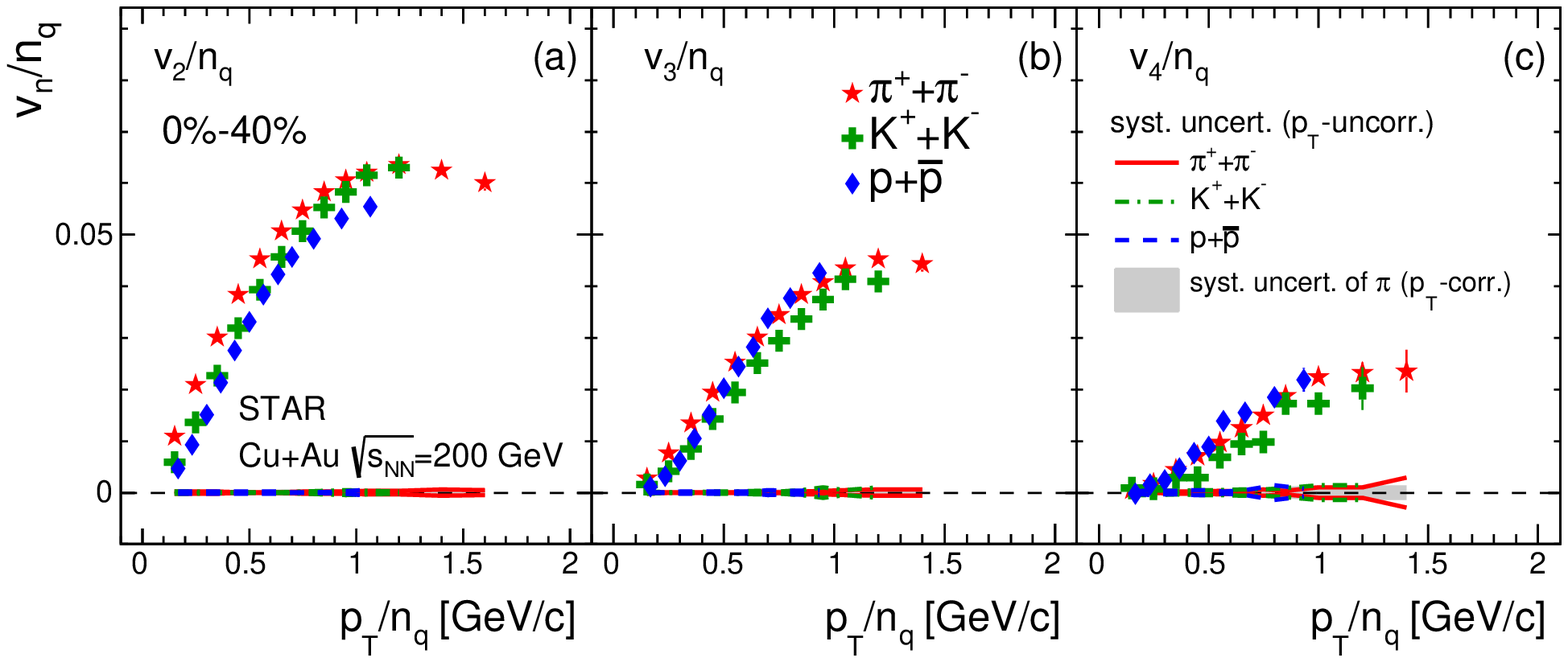}
\end{center}
\caption{(Color online) NCQ scaling of $v_2$, $v_3$, and $v_4$ of
  $\pi^{+}+\pi^{-}$, $K^{+}+K^{-}$, and $p+\bar{p}$ as a function of
  $\pt/n_q$ in the 0\%-40\% centrality bin.
  Solid lines represent \pt-uncorrelated systematic uncertainties for each species.
Shaded bands represent \pt-correlated systematic uncertainties for pions.}
\label{fig:pidvn_ncqpt}
\end{figure*}

The baryon-meson splitting in the flow coefficients was already
observed in symmetric collisions and indicates the collective flow at
a partonic level, which can be tested by the number of constituent
quark (NCQ) scaling.  The idea of the NCQ scaling is based on the
quark coalescence picture of hadron production in intermediate
\pt~\cite{Voloshin:2002wa,Molnar:2003ff}.  In this process, hadrons at
a given \pt are formed by $n_q$ quarks with transverse momentum
$\pt/n_q$, where $n_q=2$ (3) for mesons (baryons).
Figure~\ref{fig:pidvn_ncqpt}(a-c) shows $v_n/n_q$ for
$\pi^{+}+\pi^{-}$, $K^{+}+K^{-}$, and $p+\bar{p}$ as a function of
$\pt/n_q$.  The scaled $v_2$, $v_3$, and $v_4$ as a function of $\pt/n_q$
seem to follow a global trend for all particles species, although
there are slight differences for each $v_n$.  For example, the pion
$v_2$ seems to deviate slightly from the other particles at low \pt region. 
This difference might be due to the effect of resonance decays or related
to the nature of pions as Goldstone
bosons~\cite{resonanceV2,Greco_2004}.  Unlike the $v_2$, kaons seem to
deviate from the other particles in $v_3$ and $v_4$.

\begin{figure*}[htb]
\begin{center}
\includegraphics[width=0.95\linewidth,trim=0 20 0 0]{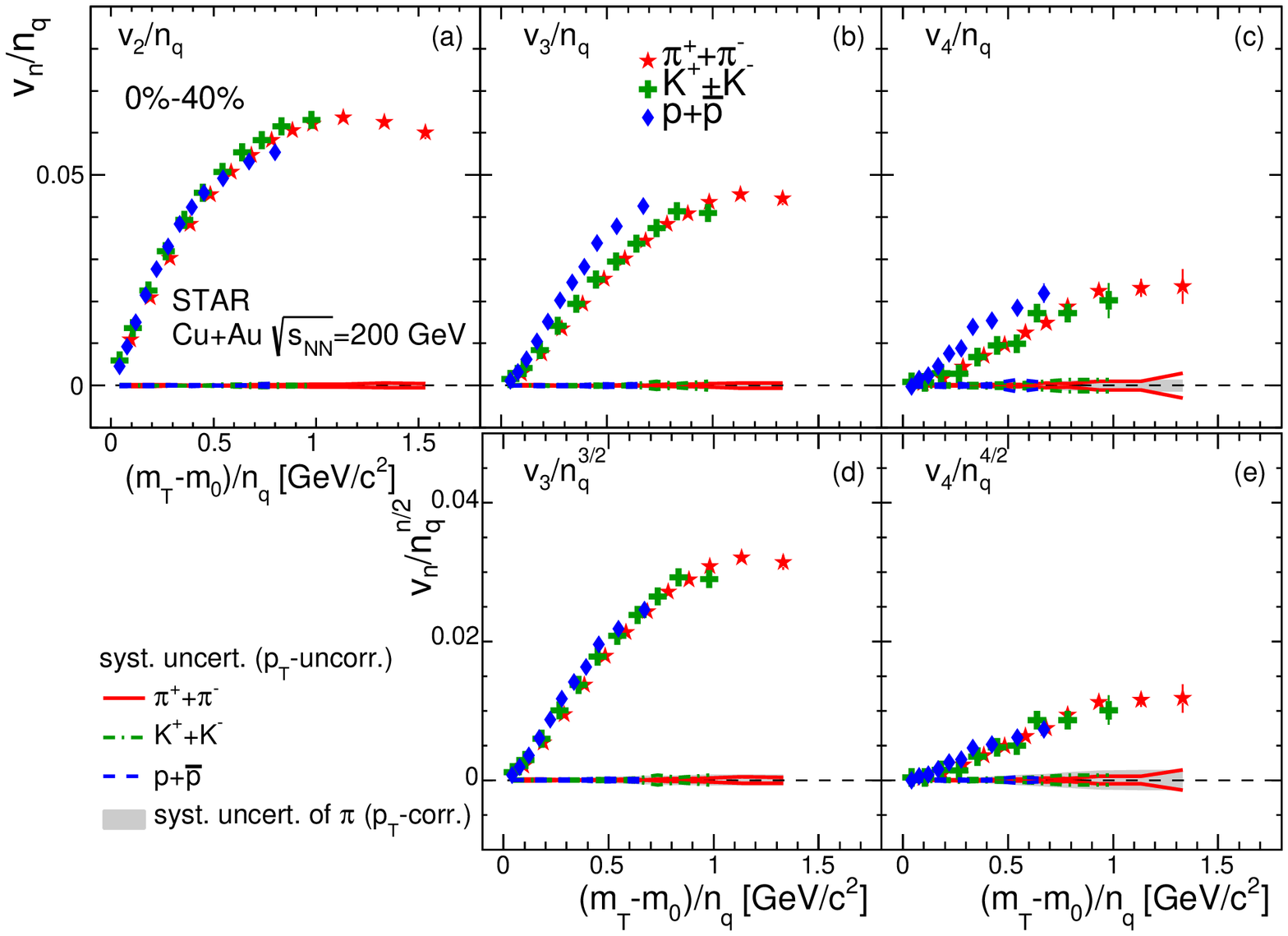}
\end{center}
\caption{(Color online) NCQ scalings of $v_2$, $v_3$, and $v_4$ of
  $\pi^{+}+\pi^{-}$, $K^{+}+K^{-}$, and $p+\bar{p}$ as a function of
  $(\mt-m_0)/n_q$ in the 0\%-40\% centrality bin.
  Solid lines represent \pt-uncorrelated systematic uncertainties for each species.
Shaded bands represent \pt-correlated systematic uncertainties for pions.}
\label{fig:pidvn_ncqket}
\end{figure*}

An empirical NCQ scaling with the transverse kinetic energy, defined as
$\mt-m_0$, is known to work well for
$v_2$~\cite{v2ket_phenix2007,v2ket_star2007}.  \mt is defined as
$\mt=\sqrt{p_T^2+m_0^2}$ and $m_0$ denotes the particle mass.  The
idea of the NCQ scaling with the transverse kinetic energy comes from an attempt
to account for the mass dependence of \pt shift during the system
radial expansion.  Figure~\ref{fig:pidvn_ncqket}(a-c) shows the NCQ
scaling with the transverse kinetic energy for $v_n$ in 0\%-40\% centrality bin.
The scaling works well for $v_2$ as reported in past studies for
symmetric collisions~\cite{v2star,v2_cucu-auau_ncq}, but it does
not work for higher harmonics.  A modified NCQ scaling for higher
harmonics, $v_n/n_{q}^{n/2}$, was proposed in Ref.~\cite{Han:2011iy}.
It works better for $v_3$ and $v_4$, as seen in
Fig.~\ref{fig:pidvn_ncqket}(d,e), as it did in Au+Au
collisions~\cite{pidvn_phenix}. Hadronic rescattering might be responsible 
for the modified scaling, but the underlying physics
is still under discussion~\cite{modNCQ,vnscaling}.

%
%

%

%

\section{Summary\label{sec:summary}}
We have presented results of azimuthal anisotropic flow measurements,
from the first- up to the fourth-order harmonics, for unidentified and
identified charged particles in Cu+Au collisions at \sqsn = 200 GeV,
as well as the directed flow of charged particles in Au+Au collisions
at \sqsn = 200 GeV from the STAR experiment.  In addition to directed
flow, the average projection of the transverse momentum on the flow
direction, \apx, was measured in the both systems.

For inclusive charged particles, the directed flow, $v_1$, was
measured as functions of $\eta$ and \pt over a wide centrality range.
The slope of the conventional $v_1(\eta)$ in Cu+Au is found to be
similar to that in Au+Au, but is shifted toward the forward rapidity
(the Cu-going direction), while the \apx in Cu+Au has a slightly
steeper slope and is shifted towards backward rapidity (the Au-going
direction). The similar slopes of $v_1$ likely indicate a similar
initial tilt of the created medium. Such a tilt seems to depend weakly
on the system size but does depend on the collision energy.  The
slight difference in slope of \apx could be explained by the momentum
balance of particles between the forward and backward rapidities and
the asymmetry in multiplicity distribution over $\eta$ in Cu+Au
collisions.  The shift of the intercept in \apx is close to the
expectation based on the shift in the center-of-mass rapidity
estimated by the number of participants in Au and Cu nuclei in a
Monte-Carlo Glauber model (Eq.~\eqref{eq:ycm}). 
Comparing slopes of $v_1(\eta)$ with those
of \apx, we conclude that in mid-central collisions the relative
contribution to conventional directed flow from the initial tilt is
about 2/3 with the rest coming from rapidity dependence of the initial
density asymmetry. 
The fluctuation component of $v_1$ in Au+Au agrees with that in Pb+Pb
collisions at \sqsn = 2.76 TeV and shows a weak centrality
dependence. This indicates that the initial dipole-like fluctuations
do not depend on the system size, the system shape (overlap region of
the nuclei), or the collisions energy.  

The mean transverse momentum projected onto the spectator plane, \apx,
shows charge dependence in Cu+Au collisions but not in Au+Au
collisions, similarly as observed in charge-dependent directed flow
reported in our previous publication~\cite{cuauv1_star}.  The observed
difference can be explained by the initial electric field due to the
charge difference in Cu and Au spectator protons.  The
charge-dependent $v_1(p_T)$ was also measured for pions, kaons, and
(anti)protons.  The pion results are very similar to our previous
results of inclusive charged particles. The charge difference of $v_1$
for kaons and protons is no larger than that of pions and consistent
with zero within larger experimental uncertainties.  These results may
indicate that the number of charges, i.e. quarks and antiquarks, at
the early time when the electric field is strong ($t<0.1$ fm/$c$) is
smaller than the number of charges in the final state.

Higher harmonic flow coefficients, $v_2$, $v_3$, and $v_4$, were also
presented as functions of \pt in various centrality bins, showing a
similar centrality dependence to those in Au+Au collisions.  The $v_2$
in Cu+Au is smaller than that in Au+Au for the same number of
participants because of different initial eccentricities.  Meanwhile,
$v_3$ scales with the number of participants between both systems,
supporting the idea that $v_3$ originates from density fluctuations in
the initial state.  For $p_T<2$ GeV/$c$, $v_2$ and $v_3$ were found to
be reasonably well reproduced by the event-by-event viscous
hydrodynamic model with the shear viscosity to entropy density
$\eta/s=0.08-0.16$ with the Glauber initial condition.  The AMPT model
calculations also qualitatively reproduced the data of $v_2$, $v_3$,
and $v_4$.

For identified particles, a particle mass dependence was observed at
low \pt for all flow coefficients ($v_1$-$v_4$), and a baryon-meson
splitting was observed at intermediate \pt for $v_2$, $v_3$, and
$v_4$, as expected from the collective behavior at the partonic level.
The number of constituent quark scaling with \pt, originating in a
naive quark coalescence model, works within $\sim 10$\% for all $v_n$.
The empirical number of constituent quark scaling with the kinetic
energy works well for elliptic flow but not for higher harmonics,
where the modified scaling works better.  This is similar to what has
been observed in Au+Au collisions. The exact reason for that is still
unknown; our new data should help in future theoretical efforts in
answering this question.

%
\begin{acknowledgments}
We thank the RHIC Operations Group and RCF at BNL, the NERSC Center at
LBNL, and the Open Science Grid consortium for providing resources and
support. This work was supported in part by the Office of Nuclear
Physics within the U.S. DOE Office of Science, the U.S. National
Science Foundation, the Ministry of Education and Science of the
Russian Federation, National Natural Science Foundation of China,
Chinese Academy of Science, the Ministry of Science and Technology of
China and the Chinese Ministry of Education, the National Research
Foundation of Korea, Czech Science Foundation and Ministry of
Education, Youth and Sports of the Czech Republic, Department of
Atomic Energy and Department of Science and Technology of the
Government of India, the National Science Centre of Poland, the
Ministry of Science, Education and Sports of the Republic of Croatia,
RosAtom of Russia and German Bundesministerium fur Bildung,
Wissenschaft, Forschung and Technologie (BMBF) and the Helmholtz
Association.
\end{acknowledgments}
\appendix*
\section{Directed flow from a ``tilted source''}\label{sec:app}

In this Appendix we derive the relation between the rapidity slopes of
$v_1$ and \apx in the ``tilted source'' scenario.
The approach used here is very similar to the one developed
in~\cite{Voloshin:1996nv}. Let us denote the invariant particle
distribution as:
\begin{eqnarray}
\frac{d^3n}{d^2p_Tdy}=J_0({p_T,y}).
\end{eqnarray}
A small ``tilt'' in $xz$ plane by an angle $\gamma$ leads to a change
in the $x$ component of the momentum $\Delta p_x=\gamma p_z = \gamma
p_T/\tan(\theta) =\gamma p_T \sinh\eta$, where $\eta$ is the
pseudorapidity.  Then the particle distribution in a tilted coordinate
system would read
\begin{eqnarray}
J &\approx& J_0+\frac{\partial J_0}{\partial p_T}\frac{\partial
  p_T}{\partial{p_x}}\Delta p_x\nonumber \\ &=& J_0\left(1 +\frac{\partial \ln
    J_0}{\partial p_T} \cos\phi \, p_T \, \gamma \, \sinh\eta \right).
\end{eqnarray}
From here one gets 
\begin{eqnarray}
v_1(p_T)=\frac{1}{2} \gamma\, p_T \sinh \eta \frac{\partial \ln
    J_0}{\partial p_T}.
\end{eqnarray}
Heavier particle spectra usually have less steep dependence on $p_T$,
which would lead to the mass dependence of $v_1(p_T)$ -- particles
with large mass would have smaller $v_1$ at a given $p_T$.
Integrating over $p_T$, and using $p_T$ weight for $\mpx$ calculation
leads to the following ratio of slopes:
\begin{eqnarray}
\frac{\displaystyle \frac{1}{\pt} \frac{d\mpx}{d\eta}}{\displaystyle
  \frac{dv_1}{d\eta}} = \frac{1}{\pt} \frac{\displaystyle \left<
  p_T^2\frac{\partial\ln J_o}{\partial p_T}\right>}{\displaystyle
  \left< p_T\frac{\partial \ln J_o}{\partial p_T}\right>}.
\end{eqnarray} 
For both the exponential form of $J_0(p_T)$ (approximately describing
the spectra of light particles) and the Gaussian form (better suited
for description of protons), this ratio equals 1.5.

\bibliography{ref_cuauvn}   
\end{document}

%% file: authorlist_11282017.tex
\affiliation{AGH University of Science and Technology, FPACS, Cracow 30-059, Poland}
\affiliation{Argonne National Laboratory, Argonne, Illinois 60439}
\affiliation{Brookhaven National Laboratory, Upton, New York 11973}
\affiliation{University of California, Berkeley, California 94720}
\affiliation{University of California, Davis, California 95616}
\affiliation{University of California, Los Angeles, California 90095}
\affiliation{Central China Normal University, Wuhan, Hubei 430079}
\affiliation{University of Illinois at Chicago, Chicago, Illinois 60607}
\affiliation{Creighton University, Omaha, Nebraska 68178}
\affiliation{Czech Technical University in Prague, FNSPE, Prague 115 19, Czech Republic}
\affiliation{Nuclear Physics Institute ASCR, Prague 250 68, Czech Republic}
\affiliation{Frankfurt Institute for Advanced Studies FIAS, Frankfurt 60438, Germany}
\affiliation{Institute of Physics, Bhubaneswar 751005, India}
\affiliation{Indiana University, Bloomington, Indiana 47408}
\affiliation{Alikhanov Institute for Theoretical and Experimental Physics, Moscow 117218, Russia}
\affiliation{University of Jammu, Jammu 180001, India}
\affiliation{Joint Institute for Nuclear Research, Dubna, 141 980, Russia}
\affiliation{Kent State University, Kent, Ohio 44242}
\affiliation{University of Kentucky, Lexington, Kentucky 40506-0055}
\affiliation{Lamar University, Physics Department, Beaumont, Texas 77710}
\affiliation{Institute of Modern Physics, Chinese Academy of Sciences, Lanzhou, Gansu 730000}
\affiliation{Lawrence Berkeley National Laboratory, Berkeley, California 94720}
\affiliation{Lehigh University, Bethlehem, Pennsylvania 18015}
\affiliation{Max-Planck-Institut fur Physik, Munich 80805, Germany}
\affiliation{Michigan State University, East Lansing, Michigan 48824}
\affiliation{National Research Nuclear University MEPhI, Moscow 115409, Russia}
\affiliation{National Institute of Science Education and Research, HBNI, Jatni 752050, India}
\affiliation{National Cheng Kung University, Tainan 70101 }
\affiliation{Ohio State University, Columbus, Ohio 43210}
\affiliation{Institute of Nuclear Physics PAN, Cracow 31-342, Poland}
\affiliation{Panjab University, Chandigarh 160014, India}
\affiliation{Pennsylvania State University, University Park, Pennsylvania 16802}
\affiliation{Institute of High Energy Physics, Protvino 142281, Russia}
\affiliation{Purdue University, West Lafayette, Indiana 47907}
\affiliation{Pusan National University, Pusan 46241, Korea}
\affiliation{Rice University, Houston, Texas 77251}
\affiliation{Rutgers University, Piscataway, New Jersey 08854}
\affiliation{Universidade de Sao Paulo, Sao Paulo, Brazil, 05314-970}
\affiliation{University of Science and Technology of China, Hefei, Anhui 230026}
\affiliation{Shandong University, Jinan, Shandong 250100}
\affiliation{Shanghai Institute of Applied Physics, Chinese Academy of Sciences, Shanghai 201800}
\affiliation{State University of New York, Stony Brook, New York 11794}
\affiliation{Temple University, Philadelphia, Pennsylvania 19122}
\affiliation{Texas A\&M University, College Station, Texas 77843}
\affiliation{University of Texas, Austin, Texas 78712}
\affiliation{University of Houston, Houston, Texas 77204}
\affiliation{Tsinghua University, Beijing 100084}
\affiliation{University of Tsukuba, Tsukuba, Ibaraki 305-8571, Japan}
\affiliation{Southern Connecticut State University, New Haven, Connecticut 06515}
\affiliation{University of California, Riverside, California 92521}
\affiliation{University of Heidelberg, Heidelberg 69120, Germany}
\affiliation{United States Naval Academy, Annapolis, Maryland 21402}
\affiliation{Valparaiso University, Valparaiso, Indiana 46383}
\affiliation{Variable Energy Cyclotron Centre, Kolkata 700064, India}
\affiliation{Warsaw University of Technology, Warsaw 00-661, Poland}
\affiliation{Wayne State University, Detroit, Michigan 48201}
\affiliation{World Laboratory for Cosmology and Particle Physics (WLCAPP), Cairo 11571, Egypt}
\affiliation{Yale University, New Haven, Connecticut 06520}

\author{L.~Adamczyk}\affiliation{AGH University of Science and Technology, FPACS, Cracow 30-059, Poland}
\author{J.~R.~Adams}\affiliation{Ohio State University, Columbus, Ohio 43210}
\author{J.~K.~Adkins}\affiliation{University of Kentucky, Lexington, Kentucky 40506-0055}
\author{G.~Agakishiev}\affiliation{Joint Institute for Nuclear Research, Dubna, 141 980, Russia}
\author{M.~M.~Aggarwal}\affiliation{Panjab University, Chandigarh 160014, India}
\author{Z.~Ahammed}\affiliation{Variable Energy Cyclotron Centre, Kolkata 700064, India}
\author{N.~N.~Ajitanand}\affiliation{State University of New York, Stony Brook, New York 11794}
\author{I.~Alekseev}\affiliation{Alikhanov Institute for Theoretical and Experimental Physics, Moscow 117218, Russia}\affiliation{National Research Nuclear University MEPhI, Moscow 115409, Russia}
\author{D.~M.~Anderson}\affiliation{Texas A\&M University, College Station, Texas 77843}
\author{R.~Aoyama}\affiliation{University of Tsukuba, Tsukuba, Ibaraki 305-8571, Japan}
% ...

\collaboration{STAR Collaboration}\noaffiliation